\documentclass[11pt]{article}
\pdfoutput=1
\usepackage[centertags]{amsmath}
\usepackage[square, comma, sort&compress,numbers]{natbib}
\usepackage{array,multirow}
\numberwithin{equation}{section}
\usepackage{amssymb,mathtools}
\usepackage{amsfonts,amsthm,stmaryrd,bm}
\usepackage{graphicx}
\usepackage{color}
\usepackage[normalem]{ulem}

\usepackage{epsfig}
\usepackage{bbold}

\usepackage{wrapfig}
\usepackage{float}
\usepackage{soul}
\usepackage{color}
\usepackage{enumitem}

\usepackage[dvipsnames]{xcolor}

\usepackage{tikz,pgf}
\usetikzlibrary{shapes}
\usetikzlibrary{calc}
\usetikzlibrary{decorations.pathmorphing}
\usetikzlibrary{decorations.pathreplacing,shapes.misc}
\usetikzlibrary{positioning}
\usetikzlibrary{arrows}
\usetikzlibrary{decorations.markings}
\usetikzlibrary{shadings}

\usetikzlibrary{intersections}

\def\sideremark#1{\ifvmode\leavevmode\fi\vadjust{\vbox to0pt{\vss
 \hbox to 0pt{\hskip\hsize\hskip1em
 \vbox{\hsize3cm\tiny\raggedright\pretolerance10000
  \noindent #1\hfill}\hss}\vbox to8pt{\vfil}\vss}}}

\def\dps{\displaystyle}

\renewcommand{\tilde}{\widetilde}
\renewcommand{\hat}{\widehat}
\newtheorem{prop}{Proposition}[section]

\newcommand{\bref}[1]{\textbf{\ref{#1}}}

\newcommand{\p}[1]{|#1|}

\renewcommand{\leq}{\,{\leqslant}\,}

\newcommand{\binner}[2]{%
  {\langle}\kern-4.15pt{\langle}#1{,}\,#2{\rangle}\kern-4.15pt{\rangle}}

\newcommand{\half}{\mathchoice{%
    \ffrac{1}{2}}{\frac{1}{2}}{\frac{1}{2}}{\frac{1}{2}}}

\newcommand{\ffrac}[2]{\raisebox{.5pt}%
  {\footnotesize$\displaystyle\frac{#1}{#2}$}\kern1pt}

\def\cA{\mathcal{A}}

\def\cC{\mathcal{C}}
\def\cD{\mathcal{D}}

\def\cH{\mathcal{H}}

\def\cL{\mathcal{L}}
\def\cM{\mathcal{M}}

\def\cQ{\mathcal{Q}}

\def\cS{\mathcal{S}}
\def\cT{\mathcal{T}}
\def\cU{\mathcal{U}}
\def\cW{\mathcal{W}}

\numberwithin{equation}{section} \makeatletter
\@addtoreset{equation}{section}

\def\be{\begin{equation}}
\def\ee{\end{equation}}
\def\ba{\begin{array}}
\def\ea{\end{array}}

\def\dps{\displaystyle}

\newdimen\tableauside\tableauside=1.0ex
\newdimen\tableaurule\tableaurule=0.4pt
\newdimen\tableaustep
\def\phantomhrule#1{\hbox{\vbox to0pt{\hrule height\tableaurule
width#1\vss}}}
\def\phantomvrule#1{\vbox{\hbox to0pt{\vrule width\tableaurule
height#1\hss}}}
\def\sqr{\vbox{%
  \phantomhrule\tableaustep

\hbox{\phantomvrule\tableaustep\kern\tableaustep\phantomvrule\tableaustep}%
  \hbox{\vbox{\phantomhrule\tableauside}\kern-\tableaurule}}}
\def\squares#1{\hbox{\count0=#1\noindent\loop\sqr
  \advance\count0 by-1 \ifnum\count0>0\repeat}}
\def\tableau#1{\vcenter{\offinterlineskip
  \tableaustep=\tableauside\advance\tableaustep by-\tableaurule
  \kern\normallineskip\hbox
    {\kern\normallineskip\vbox
      {\gettableau#1 0 }%
     \kern\normallineskip\kern\tableaurule}%
  \kern\normallineskip\kern\tableaurule}}
\def\gettableau#1 {\ifnum#1=0\let\next=\null\else
  \squares{#1}\let\next=\gettableau\fi\next}

\def\cA{\mathcal{A}}

\def\cC{\mathcal{C}}
\def\cD{\mathcal{D}}

\def\cH{\mathcal{H}}

\def\cL{\mathcal{L}}
\def\cM{\mathcal{M}}

\def\cQ{\mathcal{Q}}

\def\cS{\mathcal{S}}
\def\cT{\mathcal{T}}
\def\cU{\mathcal{U}}
\def\cV{\mathcal{V}}
\def\cW{\mathcal{W}}

\numberwithin{equation}{section} \makeatletter
\@addtoreset{equation}{section}

\def\ads2{\text{AdS}_{2}}
\def\cft1{\text{CFT}_{1}}
\def\dual{\text{AdS}_{2}/\text{CFT}_{1}}

\def\be{\begin{equation}}
\def\ee{\end{equation}}
\def\ba{\begin{array}}
\def\ea{\end{array}}

\def\dps{\displaystyle}

\def\ba{\begin{array}}
\def\ea{\end{array}}

\def\dps{\displaystyle}

\def\inplus{\oplus}

\def\hs{\mathfrak{hs}[\lambda]}

\def\p{\partial}

\def\Ypar{Z_{\parallel}}
\def\Yper{Z_\perp}

\def\tP{\mathbb{P}}
\def\tL{\mathbb{L}}
\def\univ{U\big(\mathfrak{so}(2,1)\big)}
\def\hd{\hat \cD_\lambda}
\def\cwf{{\bf \Delta}(N_{\parallel})}

\newcommand*{\bra}[1]{\langle{#1}|}
\newcommand*{\ket}[1]{|{#1}\rangle}

\usepackage{jheppub}

\makeatletter
\def\@fpheader{\vspace{-.1cm}}
\makeatother

\title{Towards higher-spin $\dual$  holography }

\author[a,b]{Konstantin\ Alkalaev}

\author[c]{Xavier\ Bekaert}

\affiliation[a]{I.E. Tamm Department of Theoretical Physics, \\P.N. Lebedev Physical
Institute,\\ Leninsky ave. 53, 119991 Moscow, Russia}
\affiliation[b]{Department of General and Applied Physics, \\
Moscow Institute of Physics and Technology, \\
Institutskiy per. 7, Dolgoprudnyi, \\141700 Moscow region, Russia}
\affiliation[c]{Institut Denis Poisson, Unit\'e Mixte de Recherche 7013,\\
Universit\'e de Tours, Universit\'e d'Orl\'eans, CNRS,\\
  Parc de Grandmont, 37200 Tours, France}

\emailAdd{alkalaev@lpi.ru}
\emailAdd{xavier.bekaert@lmpt.univ-tours.fr}

\abstract{We aim at formulating a higher-spin gravity theory around $\ads2$ relevant for holography. As a first step, we investigate  its kinematics by identifying the low-dimensional cousins of the standard higher-dimensional structures in higher-spin gravity such as the singleton, the higher-spin symmetry algebra, the higher-rank gauge and matter fields, etc. In particular, the higher-spin algebra is given here by $\hs$ and parameterized by a real parameter $\lambda$. The singleton is defined to be a Verma module of the $\ads2$ isometry subalgebra $\mathfrak{so}(2,1) \subset \hs$ with conformal weight $\Delta = \frac{1\pm\lambda}{2}\,$. On the one hand, the spectrum of local modes is determined by the Flato-Fronsdal theorem for the tensor product of two such singletons. It is given by an infinite tower of massive scalar fields in $\ads2$ with ascending  masses expressed in terms of $\lambda$. On the other hand, the higher-spin fields arising through the gauging of $\hs$ algebra do not propagate local degrees of freedom. Our analysis of the spectrum suggests that $\ads2$ higher-spin gravity is a theory of an infinite collection of massive scalars with fine-tuned masses, interacting with infinitely many topological gauge fields. Finally, we discuss the holographic $\cft1$ duals of the 
kinematical structures identified in the bulk. 
}

\arxivnumber{}

\begin{document}

\maketitle
\flushbottom

\section{Introduction and summary}
The recent surge of interest in $\dual$ correspondence prompted by the Sachdev-Ye-Kitaev (SYK) model \cite{Sachdev:1992fk,Kitaev,Maldacena:2016hyu} (see e.g. \cite{Sarosi:2017ykf,Rosenhaus:2018dtp} for reviews) motivates a thorough investigation of lowest-dimensional higher-spin (HS) holographic duality. Despite important distinctions, the analogies between the SYK model and $O(N)$ vector models in higher dimensions suggest that (broken) HS symmetries should play an important role, at least in some degenerate limits (e.g. in the quadratic case $q=2$ or in the vicinity of the ultraviolet Gaussian fixed point) of the $O(N)$-singlet sector. It is comforting that the HS extension of Jackiw-Teitelboim gravity \cite{Teitelboim:1983ux,Jackiw:1984je,Fukuyama:1985gg} is already available in the metric-like formulation, as well as in the frame-like formulation, i.e. as a BF theory for the two-dimensional HS algebra \cite{Alkalaev:2013fsa,Alkalaev:2014qpa,Grumiller:2013swa}. However, the bulk dual of SYK model should also contain a tower of matter fields dual to the single-trace (i.e. bilinear $O(N)$-singlet) 
primary operators on the boundary.

Before investigating some dynamical issues in the future, a systematic study of the lowest-dimensional analogues of the (by-now standard) kinematical foundation of HS gravity theories is in order.
For instance, the detailed study of HS asymptotic symmetries has already begun \cite{Grumiller:2013swa,Grumiller:2015vaa,Gonzalez:2018enk}. However, the analogue of the key group-theoretical results underlying HS gravity theories (such as the oscillator realizations of the HS algebra, the adjoint vs twisted-adjoint modules, etc) remain to be settled.
This is the goal of the present paper.

Let us recall that, in any dimension, the spectrum of HS gravity theories is organised in terms of the corresponding higher-spin algebra, of which three irreducible representations play an instrumental role:

\noindent$\bullet$ The \textit{singleton module} which plays the role of the fundamental representation of the HS algebra. From a holographic perspective, it is the fundamental field in the free conformal field theory (CFT) living on the boundary. In other words, the singleton module corresponds to a generalised free field of arbitrary conformal weight. In CFT$_1$, the HS algebra is a single copy of the Lie algebra $\hs$ defined in \cite{Feigin,Vasiliev:1989re}. We explain that any primary field of conformal weight $\Delta$ together with all its descendant, spanning  a Verma module ${\cal V}_\Delta$ of $\mathfrak{so}(2,1)$, can be taken as a singleton module. In fact, such a Verma module ${\cal V}_\Delta$ is known \cite{Feigin} to carry an indecomposable representation of the HS algebra $\hs$  for $\Delta=\frac{1-\lambda}2$ with $\lambda$ a real number.

\noindent$\bullet$ The \textit{adjoint module} is the higher-spin algebra itself, which (as any Lie algebra) carries the adjoint representation. From a holographic perspective, the HS algebra is the algebra of symmetries (which is a global symmetry in the boundary and which is gauged in the bulk). On the boundary, it can be defined as the Lie algebra of symmetries of the singleton. In the bulk, the ``gauge'' sector of HS gravity is described \`a la Cartan by a connection one-form taking value in the adjoint module. For instance, HS gravity on AdS$_2$ is a two-dimensional BF theory for the Lie algebra $\hs$ \cite{Alkalaev:2013fsa,Alkalaev:2014qpa}. In two and three spacetime dimensions, the gauge sector of HS gravity is topological, so one has to introduce a matter sector in order to have local degrees of freedom in the bulk.

\noindent$\bullet$ The \textit{twisted-adjoint module} in fact corresponds to the propagating sector of HS gravity. The so-called ``twist'' is an involutive automorphism of the HS algebra, corresponding intuitively to an inversion on the boundary and to a PT transformation in the bulk. In practice, it defines a twisted action of the HS algebra on itself. In the bulk, the ``matter'' sector of HS gravity is described by a covariantly-constant zero-form taking values in the twisted-adjoint module. In HS holography, this sector is dual to the spectrum of single-trace operators on the boundary. The identification of the spectrum of $\ads2$ fields sitting in the twisted-adjoint module for $\hs$ is the main result of our investigation.

As a corollary of our analysis, we claim that {\it an infinite tower of $\ads2$ scalar fields $\Phi_n$ ($n=0,1,2,3,\cdots$) with increasing masses $M_n$ determined by}
\be\label{m2R}
R^{2}_{_{\hspace{-0.5mm}\rm AdS}}M^2_n  = (n-\lambda)(n-\lambda+1)\;,
\ee
{\it forms a higher-spin multiplet corresponding to the twisted-adjoint module} (where
$R_{_{\hspace{-0.5mm}\rm AdS}}$ stands for the curvature radius of the $\ads2$ spacetime).
More precisely, these scalar fields $\Phi_n$ form a representation of the higher-spin algebra $\hs$. In particular,
the subset of scalar fields $\Phi_n$ for even (respectively, odd) integers $n$ forms an \textit{irreducible} representation.
The resulting higher-spin invariant Lagrangian on the $\ads2$ spacetime can be written schematically as 
\be
\label{lagrange}
\cL_{\lambda} = \half\sum_{n=0}^\infty \Big[(\nabla\Phi_n)^2 -  M^2_n \Phi_n^2\Big] + \sum_{m,n,p=0}^\infty g_{mnp}\,\Phi_m\Phi_n\Phi_p+...\;, 
\ee
where $\Phi_n$ are scalar fields with the above masses \eqref{m2R} fixed by the higher-spin symmetry $\hs$  
and $g_{mnp}$ denote some (yet unknown) cubic coupling constants, while the ellipses stand for interaction terms of higher order in the number of fields or derivatives.
Applying the standard $\dual$ dictionary on \eqref{m2R}, the boundary dual of this tower of scalar fields should be a collection of primary operators ${\cal O}_n$ with conformal weight $\Delta_n=n-\lambda+1$. This spectrum of primary operator matches the collection of $O(N)$-singlet operators 
\be
{\cal O}_n\sim\vec\psi\cdot\partial^n\vec\psi+\cdots\;,
\ee
bilinear in a large number $N$ of one-dimensional boundary fields $\vec\psi = \vec\psi(\tau)$ in the vector representation of $O(N)$ with conformal weight $\Delta=\frac{1-\lambda}2$. To be more precise, this matching applies only in the absence of anomalous dimensions. For instance, in the SYK model the boundary fields $\vec\psi$ are $d=1$ Majorana fermions (so the bilinear operators ${\cal O}_n$ exist only for odd $n$). At the interacting fixed point, they acquire anomalous dimensions of order one which indicates a hard breaking of HS symmetry, in contradistinction with what happens for the vector models in higher dimensions where the anomalous dimensions are typically of order $1/N$ and HS symmetries are softly broken by quantum corrections in the bulk. Nevertheless, we believe worthwhile to investigate the HS dual of some degenerate limits of SYK models with unbroken HS symmetries as a small preliminary step in the search of their mysterious bulk dual.\footnote{See e.g. \cite{Jevicki:2016bwu,Jevicki:2016ito}
for an early proposal in terms of bi-local collective fields (with some
similarities to the previous bi-local field approach to higher-dimensional
HS holography, cf. \cite{Das:2003vw}) and also \cite{Gross:2017vhb,Peng:2018zap} for discussions of possible HS holographic duals of the SYK model and its generalizations.}

The paper is organized as follows. 

In Section \bref{sec:singleton}, we first discuss singletons in arbitrary $d$ dimensions as the so-called minimal representation of $\mathfrak{so}(d,2)$ algebra and accordingly we discuss singletons in $d=1$. In general, one-dimensional singletons will be defined as Verma modules $\cV_{\Delta}$ of the $\mathfrak{so}(2,1)$ algebra with real conformal weights $\Delta$. We review the classification of all such modules in terms of their unitarity and (ir)reducibility properties, comparing with the corresponding HS algebras. In the context of higher-dimensional HS gravity, we discuss in which sense they would correspond to type-A, type-B, etc, models. The main result reviewed here is the Flato-Fronsdal type theorem which describes the tensor product of two singletons as an infinite collection of Verma modules with running conformal weights.      

In Section \bref{sec:HS_alg}, we introduce $d=1$ higher-spin algebra as the symmetry algebra of a given singleton module $\cV_{\Delta}$, where the conformal weight can be parameterized by a new parameter $\lambda$ as $\Delta = \half(1\pm \lambda)$. The resulting algebra is $\hs$, originally introduced in \cite{Feigin,Vasiliev:1989re}. 
   
In Section \bref{sec:oscillator}, the higher-spin algebra  is described using the oscillator technique and the associated star-product. We show that $\hs$ realized as the quotient of the universal enveloping algebra $\univ$ over the ideal generated by the Casimir element can be effectively described using the star-product calculation machinery.       

In Section \bref{sec:module}, we discuss $\mathfrak{so}(2,1)$ module structure of the universal enveloping algebra $\univ$ and of the higher-spin algebra $\hs$. We describe two types of their representations mentioned above: the adjoint and twisted-adjoint representations. In particular, we investigate how two representations branch with respect to the subalgebra $\mathfrak{so}(2,1)$.     

In Section \bref{sec:verma}, the previous modules are described in terms of the $\mathfrak{so}(2,1)$ covariant auxiliary variables used to formulate the higher-spin algebra $\hs$. These results allow to describe the higher-spin algebra as an infinite collection of Verma modules with running conformal dimensions and thereby identify the right-hand-side of  the Flato-Fronsdal theorem as the twisted-adjoint representation.     

In Section \bref{sec:unfolded}, we formulate the (linearized)  higher-spin dynamics on $\ads2$. It is shown that the fields are naturally divided into topological gauge fields arising through the gauging of $\hs$ algebra and matter fields propagating local degrees of freedom. Gauge and matter  fields belong, respectively, to the adjoint and twisted-adjoint representations of $\hs$ algebra. The matter sector is given by an infinite collection of massive scalars with masses explicitly matching the spectrum of the Flato-Fronsdal theorem.      


A few comments on interactions are given in Section \bref{sec:con}. Some technical points are further elaborated in the appendices. Appendix \bref{app:verma} reviews basic facts from $\mathfrak{sl}(2, \mathbb{R})$ representation theory needed to discuss Verma modules. Appendix \bref{app:star} collects  various star-product expressions used in practice. Appendices \bref{app:isomorphism} and \bref{app:isomorphism2} contain the proofs of the propositions in the main text about two convenient bases in the higher-spin algebra $\hs$. Appendix \bref{app:twisted} builds the most general class of $\mathfrak{so}(2,1)$ actions on $\univ$ realized by inhomogeneous first-order differential operators in auxiliary variables.

\section{Singletons in one dimension}
\label{sec:singleton}

\subsection{Singletons: fundamental representation}

In dimensions $d\geqslant 3$, singletons are traditionally defined as ``minimal'' (in mathematical jargon) or ``ultrashort'' (in physicist jargon) unitary irreducible $\mathfrak{so}(d,2)$-module with bounded energy (see e.g. \cite{Bekaert:2011js} and refs therein). 
In practice, these irreducible modules span only one line in the weight space of $\mathfrak{so}(d,2)$, hence their name ``singleton''.\footnote{Strictly speaking: they span two lines for the parity-invariant combination in even $d$, or a finite number of lines for their higher-order non-unitary generalisations sometimes called ``multi-linetons'' for this reason.} They also correspond to irreducible $\mathfrak{so}(d,2)$-modules which do not branch: they remain irreducible (or at most split in two irreducible submodules) when the conformal algebra $\mathfrak{so}(d,2)$ is restricted to the isometry subalgebras $\mathfrak{so}(d-1,2)$, $\mathfrak{iso}(d-1,1)$ or $\mathfrak{so}(d,1)$ \cite{Angelopoulos:1997ij}. In terms of AdS$_{d+1}$/CFT$_d$ dictionary, this translates as follows: singletons correspond to bulk fields whose dynamical degrees of freedom sit on the conformal boundary \cite{Angelopoulos:1980wg}.

As usual in a conformal
field theory, a primary field (together with its descendants) is
described by a (generalized) Verma module
${\cal V}(\Delta; \vec s)$, which is induced from the $\mathfrak{so}(2) \oplus
\mathfrak{so}(d)$ module with lowest weight $[\Delta;\vec s]$. Here, $\Delta$ is a
real number corresponding to the conformal weight, and
$\vec s:=(s_1,\dots,s_r)$ is an integral dominant $\mathfrak{so}(d)$-weight
made of $r$ ``spin'' labels, where $r$ is the rank of $\mathfrak{so}(d)$. 
The irreducible module obtained as a quotient of the Verma module
${\cal V}(\Delta; \vec s)$ by its maximal submodule will be denoted
${\cal D}(\Delta; \vec s)$.
The simplest example of singleton is the conformal scalar, also called ``Rac'', which exists in any dimension $d$. It corresponds to the irreducible quotient
\be
\mbox{Rac}(d,2):={\cal D}(\tfrac{d-2}2;{0})={\cal V}(\tfrac{d-2}2;0)\,/\,{\cal V}(\tfrac{d+2}2;{0})\;.
\ee

Things drastically simplify in $d=1$ dimension. The irreducible lowest-weight modules are entirely characterised by the conformal weight, since there is no more any ``spin'' label.
In fact, the $d=1$ conformal algebra is the smallest non-compact simple Lie algebra $$\mathfrak{so}(1,2)\cong\mathfrak{sp}(2,{\mathbb R})\cong\mathfrak{su}(1,1)\cong\mathfrak{sl}(2,{\mathbb R})\,.$$
All lowest-weight unitary irreducible $\mathfrak{so}(1,2)$-modules are reasonable candidates for the role of singleton because none of them is ``shorter'' or ``smaller'' than the others. These modules belong to the so-called discrete series of $\mathfrak{so}(1,2)$. We will not use this terminology here because it might bring confusion since it contains a \textit{continuum} of unitary irreducible representations.
Up to normalization, the corresponding mass-like term of such $AdS_2$ fields in the bulk is given by the eigenvalue $m^2$ of the quadratic Casimir operator ${\cal C}_2\big(\mathfrak{so}(1,2)\big)$,
\be
\label{m2}
m^2 = \frac{1}{4}(\lambda^2-1) = - \Delta_+ \Delta_-\;,
\ee
where $\lambda$ is a real parameter which, in this section, will be assumed to be non-negative ($\lambda\geqslant 0$) without loss of generality, and
\be
\label{delta_pm}
\Delta_\pm = \half(1\pm\lambda)
\ee
are the conformal weights corresponding to the bulk field with either Dirichelet ($\Delta_+$) or Neumann ($\Delta_-$) boundary condition, satisfying  $\Delta_+\geqslant\Delta_-$ and $\Delta_++\Delta_-=1$. 

\paragraph{Unitary singletons:} 
All lowest-weight Verma modules ${\cal V}_\Delta$ of $\mathfrak{so}(1,2)$ are candidate singletons in $d=1$ dimensions.  They are unitary iff the conformal weight is non-negative,\footnote{The proof is reviewed in Appendix \bref{app:verma}.} 
\be
{\cal V}_\Delta\quad\mbox{unitary}\quad\Longleftrightarrow\quad\Delta\geqslant0\;.
\ee
This is equivalent to the Breitenlohner-Freedman bound \cite{Breitenlohner:1982jf} on the mass-square, $m^2\geqslant-\tfrac14$.
Moreover, these modules are irreducible iff the weight is positive, 
\be
{\cal V}_\Delta\quad\mbox{unitary and irreducible}\quad\Longleftrightarrow\quad\Delta>0\;.
\ee 
Accordingly, this known\footnote{See e.g. \cite{Kitaev:2017hnr} for an introduction to the representation theory of the Lie group $SL(2,{\mathbb R})$. The results for the Lie algebra $\mathfrak{sl}(2,{\mathbb R})$ are easily extracted by relaxing restrictions on weights coming from the global structure of the group. In some sense, one allows for ``infinite-valued'' representations of the group (not only single- or double-valued).} classification of lowest-weight $\mathfrak{so}(1,2)$-modules leads to three relevant cases:
 
\noindent$\bullet$ \textit{Holographic degeneracy:} The unitarity region $0\leqslant\lambda<1$ is exceptional because, in the previous sense, both values $\Delta_\pm={\frac{1\pm\lambda}2}$ of the conformal weight (of the primary field and of its shadow) are unitary. This corresponds to the phenomenon of ``holographic degeneracy'' in the AdS/CFT context \cite{Klebanov:1999tb} where the $AdS_2$ scalar field has negative mass-square $-\tfrac14\leq m^2<0$, cf \eqref{m2}, but is unitary since it is above the Breitenlohner-Freedman bound. We will refer to the continuum of possible higher-spin algebras and theories based on those free unitary singletons with $0<\lambda<1$ as ``type-$\lambda$''.\footnote{Note that $0\leqslant\lambda\leqslant 1$ is the usual range of values of $\lambda$ considered in Gaberdiel-Gopakumar $AdS_3/CFT_2$ higher-spin holography (cf. \cite{Gaberdiel:2010pz} and sequel) where there are two copies of $\hs$.}

\noindent$\bullet$ \textit{Spinor singleton:} The limiting case $\lambda=1$ (equivalently, $m^2=0$) is special because $\Delta_-=0$ is the conformal weight of the trivial representation and $\Delta_+=1$ is conformal weight of its invariant subspace. Physically, a primary field with $\Delta=0$ corresponds to an off-shell $d=1$ spinor singleton ${\cal V}_{0}$ which is a reducible $\mathfrak{so}(1,2)$-module since it decomposes into the semidirect sum ${\cal V}_{0}\cong{\cal D}_0\niplus{\cal D}_1$ of irreducible modules. In fact, the on-shell $d=1$ spinor, ${\cal D}_0={\cal V}_0/{\cal V}_1$, is spanned by the constants, i.e. it corresponds to the trivial representation. The higher-spin theory corresponding to the off-shell spinor without the zero-mode, ${\cal D}_1={\cal V}_1$, will be called the ``type-B'' theory by analogy with the terminology for the holographic dual of the free spinor in higher dimensions (see \cite{Sezgin:2003pt,Alkalaev:2012rg}). 

\noindent$\bullet$ \textit{Singletons of any spin:} The unitarity region $\lambda>1$ corresponds to the generic case where only the conformal weight $\Delta_+={\frac{1+\lambda}2}$ is admissible.
All those Verma modules ${\cal V}_{\frac{1+\lambda}2}$ are irreducible and unitary with positive mass-square $m^2>0$. In some sense, they can be thought as (off-shell) spin-$j$ singletons. Accordingly, we will refer to the corresponding theories as ``type-J'' (cf. \cite{Basile:2018dzi} for this terminology) when $\lambda=N\in \mathbb N$, by analogy with singletons in higher dimensions where spin is quantised in general.

\paragraph{Non-unitary singletons:} For positive integer $\lambda=N>1$ (with $N\in\mathbb N$), the $\mathfrak{so}(1,2)$-module ${\cal V}_{\frac{1-N}2}$ is reducible, with ${\cal V}_{\frac{1+N}2}$ as (unitarisable) submodule. The corresponding quotient is the (non-unitarisable) irreducible module 
\be
\label{finite-dim}
{\cal D}_{\frac{N-1}2}\cong{\cal V}_{\frac{1-N}2}\,/\,{\cal V}_{\frac{1+N}2}
\ee
of finite dimension $N$. We will refer to this countably-infinite collection of possible theories based on such finite-dimensional non-unitary representations as ``type-N''.\footnote{See recent discussion of finite-dimensional singleton modules and associated topological HS models in \cite{Grigoriev:2019xmp}.} Note that once one drops the unitarity condition, one has higher-order versions of singletons in higher dimensions as well, for instance the $\mathfrak{so}(d,2)$-modules \cite{Basile:2014wua}
\be
\mbox{Di}_\ell:=\frac{{\cal V}\left(\frac{d+1}2-\ell;{\bf \frac12}\right)}{{\cal V}\left(\frac{d-1}2+\ell;{\bf \frac12}\right)}\,,\quad \mbox{Rac}_\ell:=\frac{{\cal V}\left(\frac{d}2-\ell;{0}\right)}{{\cal V}\left(\frac{d}2+\ell;{0}\right)}\,.
\ee
The theories based on free Rac$_\ell$ (respectively, Di$_\ell$) fields are nowadays called ``type-A$_\ell$'' (respectively, ``type-B$_\ell$'').
The Rac$_\ell$ (respectively, Di$_\ell$) are irreducible for positive lowest-energy $E_0=\Delta>0$ but become reducible for even (respectively, odd) $d$ and negative lowest-energy, in which case the irreducible modules are actually finite-dimensional, e.g. ${\cal D}(-n,0)$ is isomorphic to the $\mathfrak{so}(d,2)$-module ${\cal D}_n$ spanned by rank-$n$ traceless symmetric tensors (i.e. spherical harmonics). Such CFTs of type-N have been discussed in \cite{Brust:2016gjy}.

\vspace{1mm}
To conclude, let us note that descending in two dimensions the discrete set of higher-dimensional higher-spin gravity theories of A,B,C,... types coalesces into a one-parametric family of higher-spin theories parameterized by the continuous parameter $\lambda$ related to the weight of the singleton module \eqref{delta_pm}.

\subsection{Spectra of bulk fields/boundary operators: bi-fundamental representation}

The spectrum of boundary ``single-trace'' operators for vector models is extremely simple since singlets of the vector representation are generated from bilinear singlets. Therefore, the spectra are determined by the decomposition of the tensor product of two singletons into irreducible $\mathfrak{so}(d,2)$-modules. This type of decomposition is nowadays referred to as a ``Flato-Fronsdal theorem'' since it was first obtained in the $d=3$ case in \cite{Flato:1978qz}. For the scalar singleton in any dimension $d$, this reads \cite{Vasiliev:2004cm,Dolan:2005wy}
\be\label{FFanyd}
\mbox{Rac}(d,2)\otimes\mbox{Rac}(d,2)=\bigoplus\limits_{s=0}^\infty{\cal D}(s+d-2;s)
\ee
where ${\cal D}(s+d-2;s)$ corresponds to the module spanned by bulk gauge fields of spin $s$, dual to conserved currents on the boundary \cite{Vasiliev:2004cm,Dolan:2005wy}.
From the bulk point of view, the modules appearing in the decomposition into irreducible $\mathfrak{so}(d,2)$-modules span the spectrum of fields in the bulk higher-spin gravity dual to the free conformal scalar. 

A standard result for the tensor product between two Verma modules of $\mathfrak{so}(1,2)$ is the  decomposition \`a la Clebsh-Gordan (see e.g. \cite{Basile:2018dzi})
\begin{equation}
\label{d=1FF!}
  {\cal V}_{\Delta_1}\otimes{\cal V}_{\Delta_2} =
  \bigoplus_{n=0}^{\infty} {\cal V}_{\Delta_1+\Delta_2+n}\,.
\end{equation}
For $\Delta_1 = \Delta_2\equiv \Delta$, one obtains the one-dimensional version of the Flato-Fronsdal theorem \eqref{FFanyd}  
\be
\label{tensorFF}
  {\cal V}_{\Delta}\otimes{\cal V}_{\Delta} =
  \bigoplus_{n=0}^{\infty} {\cal V}_{2\Delta+n}\,.
\ee 
Substituting $\Delta = \Delta_\pm=\half(1\pm\lambda)$ yields 
\begin{equation}
\label{tensor_FF}
  {\cal V}_{\tfrac{1\pm\lambda}2}\otimes{\cal V}_{\tfrac{1\pm\lambda}2} =
  \bigoplus_{n=0}^{\infty} {\cal V}_{n+1\pm\lambda}\,.
\end{equation}
The spectrum of conformal weights on the right-hand-side is
\be
\label{dim}
\Delta^\pm_n = n+1\pm \lambda \;,\qquad
n = 0,1,2,...\;.
\ee
The corresponding spectrum of mass-squares reads
\be
\label{mass_squared}
(m^\pm_n)^2 =  \Delta^\pm_n (\Delta^\pm_n-1) = (n\pm\lambda + 1)(n\pm\lambda)\;,
\ee
which reproduces \eqref{m2R} by inserting the curvature radius and taking the minus sign. Let us stress that, for notational simplicity,  we will assume in the rest of the paper that $\lambda$ can take any real value (positive, zero, or negative). For the sake of definiteness, we set by default $\Delta_- = \frac{1-\lambda}{2}$ as the conformal weight of the singleton, which is allowed if one forgets about unitarity issues (in fact, they will not be addressed any more in the sequel since unitarity has been discussed in details in this previous subsection).

The spectrum of the type-N theory for the finite-dimensional ``singleton'' module \eqref{finite-dim} can be obtained from the usual Clebsh-Gordan decomposition of the tensor product of two finite-dimensional modules
\be
\label{tensor}
  {\cal D}_{\frac{N-1}2}\otimes{\cal D}_{\frac{N-1}2} =
  \bigoplus_{n=0}^{N-1} {\cal D}_{j}\,.
\ee 

As a conclusion to this section, let us note that formula \eqref{tensor_FF} has a natural interpretation in AdS$_3$/CFT$_2$ holography as well, due to the isomorphism $\mathfrak{so}(2,2)=\mathfrak{so}(2,1)\oplus\mathfrak{so}(2,1)$. A massive scalar on AdS$_3$ of lowest energy $E_0=1\pm\lambda$ is described by an irreducible $\mathfrak{so}(2,2)$-module given by the left-hand-side in \eqref{tensorFF}.\footnote{See e.g. Section 4 in \cite{Basile:2018dzi} for a review of the group-theoretical dictionary.}
The index $n$ appearing in the decomposition with respect to the subalgebra $\mathfrak{so}(2,1)\subset\mathfrak{so}(2,2)$ in the right-hand-side of \eqref{tensorFF} can in fact be interpreted as labeling the Fourier modes in a Kaluza-Klein-like compactification \cite{Metsaev:2000qb,Artsukevich:2008vy,Gross:2017aos}. It suggests that the integer-spaced spectrum \eqref{dim} of  massive scalar modes in the HS theory in two dimensions can be obtained from the dimensionally reduced HS theory of Prokushkin and Vasiliev in three dimensions, where all local degrees of freedom are propagated by a single scalar field of mass $M^2 = 1-\lambda^2$ \cite{Prokushkin:1998bq}.\footnote{Recall that the HS symmetry algebra in three dimensions is given by two copies of the HS symmetry algebra in two dimensions, ${\mathfrak g}_\lambda = \hs\oplus \hs$ (cf. Section \bref{sec:HS_alg}), so that the parameter $\lambda$ defines the spectrum in both cases.}      

\section{Higher-spin algebra in two dimensions}
\label{sec:HS_alg}

The modern definition of higher-spin algebras is as Lie algebras of endomorphisms of the singleton (see e.g. \cite{Konstein:1989ij,Fradkin:1990ki,Fradkin:1990ir} for some early seminal works and \cite{Boulanger:2013zza,Joung:2014qya,Fernando:2015tiu} for some recent systematic discussions of higher-spin algebras in any dimension).

\subsection{Definition}

Let us denote by ${\cal U}(\mathfrak{g})$ the universal enveloping algebra of the Lie algebra $\mathfrak g$. 
The annihilator Ann$V\subset{\cal U}(\mathfrak{g})$ of an irreducible $\mathfrak g$-module $V$ is called a primitive ideal. Due to Schur lemma, a primitive ideal contains in particular the ideal generated by the Casimir operators of $\mathfrak{g}$ minus their eigenvalue on $V$. 
The Lie algebra $\mathfrak{gl}$(V) of endomorphisms of an irreducible $\mathfrak g$-module $V$ can be defined as the realisation of ${\cal U}(\mathfrak{g})$ on $V$: it is isomorphic to the quotient ${\cal U}(\mathfrak{g})/$Ann$(V)$ of the universal enveloping algebra by the primitive ideal, endowed with the commutator as Lie bracket. If the $\mathfrak g$-module $V$ is unitary, then one will (implictly) consider the related real form of the latter algebra. One may remove the Abelian ideal $\mathfrak{u}(1)\cong \mathbb R$ part proportional to the unity in order to obtain a simple algebra (this is the standard convention in $d=1,2$ but not for $d>2$).

The higher-spin algebra of the singleton ${\cal V}_{\frac{1\pm\lambda}2}$ is the quotient
\be
\mathfrak{gl}\big({\cal V}_{\frac{1\pm\lambda}2}\big)=\frac{{\cal U}\big(\mathfrak{so}(1,2)\big)}{\mbox{Ann}({\cal V}_{\frac{1\pm\lambda}2})}\cong\mathfrak{gl}[\lambda]
\ee
isomorphic to the algebra $\mathfrak{gl}[\lambda]\cong\mathfrak{gl}[-\lambda]$ defined in \cite{Feigin} as the quotient
\be
\label{quotient}
\mathfrak{gl}[\lambda]\, \coloneqq \,\frac{{\cal U}\big(\mathfrak{so}(1,2)\big)}{\Big({\cal C}_2\big(\mathfrak{so}(1,2)\big)-\frac14(\lambda^2-1)\Big){\cal U}\big(\mathfrak{so}(1,2)\big)}\;,
\ee
of the universal enveloping algebra of $\mathfrak{sl}(2,{\mathbb R})\cong
\mathfrak{so}(1,2)$ by the ideal generated by the eigenvalue $\frac14(\lambda^2-1)$ of the quadratic Casimir element ${\cal C}_2\big(\mathfrak{so}(1,2)\big)\in{\cal U}\big(\mathfrak{so}(1,2)\big)$. 
The corresponding higher-spin algebra, nowadays often denoted $\mathfrak{hs}[\lambda]$, is traditionnaly defined by subtracting the one-dimensional Abelian ideal:
\be
\mathfrak{gl}[\lambda]=\mathfrak{u}(1)\oplus\mathfrak{hs}[\lambda]\,.
\ee
In this way, the higher-spin algebra $\mathfrak{hs}[\lambda]$ is simple for $|\lambda|\notin\mathbb N$. 
Two algebras $\mathfrak{gl}[\lambda_1]$ and $\mathfrak{gl}[\lambda_2]$ are isomorphic iff $\lambda_1=\pm\lambda_2$. Accordingly, the singleton ${\cal V}_{\frac{1\pm\lambda}2}$
and its shadow ${\cal V}_{\frac{1\mp\lambda}2}$ define the same higher-spin algebra $\mathfrak{hs}[\lambda]$.

The exceptional isomorphism $\mathfrak{so}(1,2)\cong\mathfrak{sp}(2,{\mathbb R})$ is such that the Verma module ${\cal V}_{\tfrac14}$ (respectively, ${\cal V}_{\tfrac34}$) is isomorphic to the subspace of even (respectively, odd) polynomials in the creation operators inside the Fock space built from one oscillator. 
Accordingly, the higher-spin algebra $\mathfrak{hs}[\tfrac12]$ is isomorphic to the subalgebra ${\cal A}^{\mbox{even}}_2$ spanned by even polynomials of the Weyl algebra ${\cal A}_2$ of two-dimensional phase space. More generally, the Verma modules ${\cal V}_{\frac{1\pm\lambda}2}$ admit a realisation in terms of deformed oscillators with deformation parameter $\nu$ defined by $\lambda=\tfrac12(\nu+1)$ \cite{Vasiliev:1989re}.
 
\subsection{Extra factorization}
\label{sec:extra}

At $\lambda=N\in \mathbb{N}$, the type-J higher spin algebra $\mathfrak{hs}[N]$ is not simple and contains an infinite-dimensional ideal to be factored out \cite{Feigin,Vasiliev:1989re}.
This infinite-dimensional ideal ${\cal J}_N\subset \mathfrak{hs}[N]$ is related to the existence of the submodule ${\cal V}_{\frac{1+N}2}\subset{\cal V}_{\frac{1-N}2}$ (see e.g. Section 4.2 of \cite{Basile:2018dzi} for some discussion). The corresponding quotient
\be
\label{integer_f}
\mathfrak{sl}(N,{\mathbb R})\cong\frac{\mathfrak{hs}[N]}{{\cal J}_N}
\ee
is the type-N higher-spin algebra $\mathfrak{sl}(N,{\mathbb R})$, which is finite-dimensional. The Lie algebra isomorphism
\eqref{integer_f} is a consequence of the module isomorphisms \eqref{finite-dim} and 
\be
\mathfrak{gl}(N,{\mathbb R})\cong\mathfrak{gl}\big({\cal D}_{\frac{N-1}2}\big)\,.
\ee
Another way to understand this result is to use the principal embedding $\mathfrak{sl}(2,\mathbb{R})\subset \mathfrak{sl}(N,\mathbb{R})$. In fact, one can decompose the reducible $\mathfrak{sl}(2,\mathbb{R})$-module $\mathfrak{sl}(N,\mathbb{R})$ into a sum of irreducible $\mathfrak{sl}(2,\mathbb{R})$-modules
\be
\label{gl(N)}
\mathfrak{sl}(N,\mathbb{R})  = \bigoplus_{j=1}^{N-1} \cD_j\;,
\ee 
where $\cD_{j}$ denotes the ``spin-$j$'' finite-dimensional $\mathfrak{sl}(2,\mathbb{R})$-irreps of dimension $2j+1$.
This is to be compared with the decomposition of the reducible $\mathfrak{sl}(2,\mathbb{R})$-module $\mathfrak{hs}[N]$ into the sum
\be
\label{hs(N)}
\mathfrak{hs}[N] = \bigoplus_{j=1}^\infty \cD_j\;.
\ee 
The ideal ${\cal J}_N\subset \mathfrak{hs}[N]$ decomposes as 
\be
\label{ideal}
{\cal J}_N = \bigoplus_{j=N}^\infty \cD_j\;,
\ee  
thereby leading to the quotient \eqref{integer_f}.

\section{Oscillator realization}
\label{sec:oscillator}

Following the original definition of bosonic higher-spin algebras in any dimension  \cite{Vasiliev:2003ev,Vasiliev:2004cm}, in two dimensions  we introduce variables $Y^A_\alpha$, with $\mathfrak{sp}(2,\mathbb{R})$-vector
index $\alpha$ and $\mathfrak{so}(2,1)$-vector index $A$,\footnote{Symplectic $\mathfrak{sp}(2,\mathbb{R})$-indices are
$\alpha,\beta,\gamma,... = 1,2$ are raised and lowered as follows: $
T^\alpha = \epsilon^{\alpha\beta}T_\beta$ and $T_\alpha = T^\beta\epsilon_{\beta\alpha}$, via the $\mathfrak{sp}(2,\mathbb{R})$ invariant metric, which is the Levi-Civita symbol
$\epsilon_{\alpha\beta} =  - \epsilon_{\beta\alpha}$ with $\epsilon_{12} =1$. Vector $\mathfrak{so}(2,1)$-indices $A,B,C, ... = 0,1,2$ are raised and lowered via the $\mathfrak{so}(2,1)$-invariant metric $\eta_{AB} =\mbox{diag}(-++)$. Lorentz vector $\mathfrak{so}(1,1)$-indices read $a,b,c, ... = 0,1$ and the $\mathfrak{so}(1,1)$-invariant metric is $\eta_{ab} = \mbox{diag}(-+)$. The  $\mathfrak{so}(2,1)$-covariant Levi-Civita symbol $\epsilon_{ABC}$ is defined by $\epsilon_{012} =1$. It satisfies the identity 
$\epsilon_{ABC}\epsilon^{MNC} = -(\delta_A^M\delta_B^N-\delta_A^N\delta_B^M)$, 
where $\epsilon^{ABC}$ is the contravariant symbol, $\epsilon^{012} =-1$. The star-(anti)commutator is denoted $\big[\,\,, \,\,\big]_*$
(respectively: $\big\{\,\,, \,\,\big\}_*$).} obeying to the commutation relations
\be
\big[Y^A_\alpha, Y^B_\beta\big]_*  = \epsilon_{\alpha\beta}\,\eta^{AB}\;.
\ee
The space of polynomials in $Y$ variables endowed with the Weyl star-product
\be\label{star}
*\,=\,\exp\left(\frac12\,\epsilon_{\alpha\beta}\,\eta^{AB}\frac{\overleftarrow{\partial}}{\partial Y^A_\alpha}\,\frac{\overrightarrow{\partial}}{\partial Y^B_\beta}\right)\,,
\ee
is  the
Weyl algebra $\cA_{3}$ which is a bi-module over $\mathfrak{so}(2,1)$ and $\mathfrak{sp}(2,\mathbb{R})$ algebras. The basis elements of the latter algebras are given by 
\be
\label{basis}
T^{AB} = \frac{1}{4}\epsilon^{\alpha\beta}\big\{ Y_\alpha^A, Y_\beta^B\big\}_*=\frac{1}{2}\epsilon^{\alpha\beta} Y_\alpha^A Y_\beta^B\;,
\qquad
t_{\alpha\beta} = \frac{1}{2}\eta_{AB}\big\{ Y_\alpha^A, Y_\beta^B\big\}_*=\eta_{AB} Y_\alpha^A Y_\beta^B\;.
\ee
The bilinears  $T^{AB}$ and $t_{\alpha\beta}$ commute with each other, $[T^{AB},t_{\alpha\beta}]_* = 0$. More precisely, the two algebras form the Howe dual pair $\mathfrak{so}(2,1)-\mathfrak{sp}(2,\mathbb{R})$ \cite{Howe}.  
 
\subsection{Universal enveloping algebra in Howe dual variables}

Let us consider the subalgebra $\cS_{3} \subset \cA_{3}$ of polynomials generated  by $\mathfrak{sp}(2,\mathbb{R})$-invariant elements, i.e.
\be
\label{spINV}
\big[t_{\alpha\beta}, F(Y)\big]_*=0\;.
\ee
We notice that since indices $A, B, ...$ run over just three values it is possible to introduce new variables
\be
\label{Tdual}
Z_A := \frac12\epsilon_{ABC}\, \epsilon^{\alpha\beta}\, Y^B_\alpha\, Y^C_\beta\;,
\ee
which are, in fact, the Hodge dualized $\mathfrak{so}(2,1)$ basis elements  $T^{AB}$ in \eqref{basis}. Hence, they satisfy the commutation relations 
\be\label{starcomZ}
\big[Z_A, Z_B\big]_* =\epsilon_{ABC} Z^C\;.
\ee

One can show that any element $F\in \cS_3$, i.e. an $\mathfrak{sp}(2,\mathbb{R})$-singlet \eqref{spINV}, can be rewritten equivalently as a power series in the $Z$ variables,
\be
\label{FZ}
F(Z) = \sum_{k=0}^\infty F_{A_1 ... A_k} Z^{A_1} ... Z^{A_k}\;,
\ee  
where the totally-symmetric $\mathfrak{so}(2,1)$-tensors $F_{A_1 ... A_k}$ are constant coefficients.
Thus, we conclude that the associative algebra $\cS_3$ of $\mathfrak{sp}(2,\mathbb{R})$-singlets and the universal enveloping algebra $\univ$ are isomorphic \cite{Alkalaev:2014qpa},
\be
\label{iso}
\cS_3 \,\cong\, \univ\;.
\ee
In fact, this is the oscillator realization of the well-known Poincar\'e-Birkhoff-Witt theorem which states that the universal enveloping algebra and the symmetric algebra are isomorphic as vector spaces. In our case, it  allows merging the standard star-product calculation technique with the abstract definition of the higher-spin algebra $\hs$ as the quotient algebra \eqref{quotient}.

\subsection{Higher-spin quotient algebra}

The higher-spin algebra $\hs$ can be defined as the quotient algebra  \eqref{quotient}. Within the oscillator realization developed in the previous sections the $\hs$ algebra elements are equivalence classes
\be
\label{equivalence}
F \sim F + \hd G\;, 
\qquad
F, G \,\in\, 
\cS_3\;,
\ee
where the equivalence relation reads 
\be
\label{hd}
\hd G \equiv \left(C_2 - \mu_\lambda\right)*G\;, 
\qquad
\mu_\lambda \coloneqq \frac{1}{4}(\lambda-1)(\lambda+1)\;,
\ee 
where the  Casimir element of $\mathfrak{so}(2,1)$ is $C_2 = \half\, T_{AB}*T^{AB}$. 
From \eqref{TAB} we can find that $ C_2 =  -\left(Z_AZ^A  - 6\right)/16$
and  its action on the universal enveloping algebra reads
\be
\label{C2F}
C_2 * F(Z) = \frac{1}{16}\left(Z^2  - (N+2)(N+3)\right)\left(\Box-1\right)F(Z)\;,
\ee
where 
\be
\label{NZB}
N= Z^A \frac{\p}{\p Z^A}\;,
\qquad
Z^2 = Z^A Z_A\;,
\qquad 
\Box = \eta^{AB}\frac{\p^2}{\p Z^A \p Z^B}\;.
\ee
Operators $N$ and $Z^2,\, \Box$ act on the expansion coefficients of the power series \eqref{FZ} as the counting and trace creation/annihilation operations, respectively. They satisfy the $\mathfrak{sl}(2,\mathbb{R})$  algebra commutation relations
\be
\label{sl2}
[\Box, N] = 2\,\Box\;,
\qquad
[Z^2, N] = -2Z^2\;,
\qquad
[\Box, Z^2] = 4N\;.
\ee

Using \eqref{C2F} we find   
\be
\label{DF}
\hd G(Z) = \left[- \mu_\lambda +\frac{1}{16}\left(Z^2  - (N+2)(N+3)\right)\left(\Box-1\right)\right ]G(Z)\;.
\ee

Let us characterize the underlying space of the higher-spin algebra. The freedom in choosing representatives of the quotient leads to the problem of which basis (i.e. particular section of the quotient) might be most appropriate for concrete computations. In what follows we take advantage of  two distinguished bases and the following proposition describes representatives that can be called {\it canonical} as they are most natural in terms of the oscillator approach used here.  

\begin{prop}
\label{prop:quotient}
As a linear space, the quotient algebra \eqref{quotient} is isomorphic to the infinite-dimensional subspace $\cH \subset \cS_3$ singled out by the constraint $\Box H(Z) = 0$. This subspace is graded by the homogeneity degree in the variables $Z^A$,
\be
\label{HS_space}
\cH = \bigoplus_{n=0}^\infty \cD_n\;, 
\qquad\quad
\cD_n = \left\{H\in\cH\,:\;\,  \Box H = 0\;, \;\; (N-n)H = 0\right\}\;.
\ee  

\end{prop}
\noindent In components, the canonical section  specified by the constraints \eqref{HS_space} is given by power series  
\be
\label{HZ}
H(Z) = \sum_{n\geqslant 0} H_{A_1 ... A_n} Z^{A_1} ... \,Z^{A_n}\;,
\qquad\;\;
\eta^{BC}H_{BCA_3 ... A_n} = 0\;. 
\ee
In other words, the canonical basis of $\hs$ is given by rank-$n$ traceless $\mathfrak{so}(2,1)$-tensors ($n=0,1,2,3,...$).

The proof is given in Appendix \bref{app:isomorphism}. It relies on the trace decompositions of elements $F\in \cS_3$ that in terms of the star-product can be represented as 
\be
\label{factoring}
F = \sum_{k\geqslant 0} \left(C_2  - \mu_\lambda\right)_*^k * F^{(k)}\;, 
\qquad
\text{where}
\qquad
\Box F^{(k)} = 0\;, 
\ee
and $\left(C_2 -\mu_\lambda\right)_*^k$ stands for the $k$-th power $*$-product.  Here, elements $F^{(k)}$ are traceless but have no definite degree. Thus, any element decomposes into a sum of a traceless element and  terms proportional to $C_2-\mu_\lambda$, i.e.  $F(Z) = F^{(0)}(Z) + \hd *G(Z)$, cf. \eqref{hd}, 
so that the factorization is now explicit,  $F^{(0)}(Z)$ is of the same form as $H(Z)$ in \eqref{HZ}. 

\section{Module structure of the universal enveloping algebra}
\label{sec:module}

There are two types of $\mathfrak{so}(2,1)$ representations that can be naturally realized on the universal enveloping  algebra $\univ$. These are defined through the  {\it adjoint} and  {\it twisted-adjoint} actions of the universal enveloping  algebra $\univ$ on itself,
followed by a restriction of the action to the subalgebra $\mathfrak{so}(2,1)\subset \univ$.
By construction, these two modules are infinite-dimensional. Moreover, they are highly reducible representations of $\mathfrak{so}(2,1)$. Depending on the type of the representation, we introduce  appropriate irreducibility conditions and decompose the (twisted-)adjoint representation into an infinite collection of irreducible (in)finite-dimensional submodules.

\subsection{Adjoint action} 
\label{sec:adjoint}

The adjoint action of the $\mathfrak{so}(2,1)$ algebra  on the universal enveloping algebra is defined by 
\be
\label{adj}
\cT_A F\, \equiv \, [T_A, F]_*=  T_{A}* F  - F*  T_{A}\;,
\qquad\;\;
\forall F\in \univ\;.
\ee
Using the star-product expressions \eqref{TAB}  we obtain that the adjoint action is given by the homogeneous differential operator 
\be
\label{adj_T}
\cT_A  = - \,\epsilon_{ABC}\, Z^B \frac{\p}{\p Z_C}\;.
\ee
The following proposition holds. 
\begin{prop}
\label{prop:adj_module}
The universal enveloping algebra $\univ$, as the adjoint $\mathfrak{so}(2,1)$-module, decomposes into an infinite collection of finite-dimensional $\mathfrak{so}(2,1)$-modules with infinite multiplicities,
\be
\label{mult}
\univ \cong \bigoplus_{m,n=0}^\infty \cD_n^{(m)}\;, 
\ee
where the label $m$ denotes $m$-th copy of module $\cD_n$ spanned by rank-$n$ traceless $\mathfrak{so}(2,1)$-tensors.
\end{prop}
\noindent The proof can be found in Appendix \bref{app:isomorphism}.

The adjoint generators \eqref{adj} have a well-defined action on the quotient algebra elements  since\footnote{Note that (anti)commutators evaluated with respect the star-product are labeled by $*$ as in, e.g. \eqref{adj}. (Anti)commutators without $*$ are defined with respect to their own multiplication laws. }
\be
\label{TAHD}
[\cT_A, \hd] = 0\;, 
\ee  
where $\hd$ is given in \eqref{DF}. From the star-product decomposition \eqref{factoring} it follows that the generators $\cT_A$ projected on the quotient algebra $\hs$ are realized by the same homogeneous differential operators \eqref{adj_T} because the trace constraint is invariant with respect to the adjoint action, 
\be\label{TAHBox}
[\cT_A, \Box]=0,
\ee
see  \eqref{Tsl2}. Indeed, acting with $\cT_A$ on an arbitrary  element $F\in \univ$ decomposed according to \eqref{factoring} we obtain $\cT_A F = \cT_A F^{(0)} + \cT_A\hd *G$. Then, using the relation \eqref{TAHD}  we see that the second term is again of the form $\hd*(\cT_AG)$. Finally, using the relation \eqref{TAHBox} we check the first term defines the action of $\mathfrak{so}(2,1)$ on the quotient algebra, since $\cT_A F^{(0)}\in \cH$.

From the above discussion together with Propositions \bref{prop:quotient} and \bref{prop:adj_module}, it follows that the underlying vector space of the higher-spin algebra $\hs$ in the canonical basis is a direct sum of an infinite number  of finite-dimensional submodules $\cD_n$, $\dim \cD_n = 2n+1$ ($n=0,1,2,...,\infty$). This is in agreement with the formula (cf. (4.16) in \cite{Basile:2018dzi}):
\be
\label{twistFF}
  \cV_{\Delta}\otimes\overline \cV_{\Delta} =
  \bigoplus_{n=0}^{\infty} \cD_{n}\,,
\ee
where the line over the Verma module stands for the contragredient representation.
Note that this formula is valid for any $\Delta$. Therefore, as $\mathfrak{so}(2,1)$-modules
the adjoint modules look isomorphic for distinct $|\lambda|$ since they decompose into the same infinite sum of irreducible modules, but as $\hs$-modules the adjoint modules are irreducible (after removing the $\mathfrak{u}(1)$ ideal) and not isomorphic for distinct $|\lambda|$.
In any case, this isomorphism is a subtle one because both sides are neither lowest nor highest weight.\footnote{Note that, although each finite-dimensional module $\cD_{n}$ in the right-hand-side is both lowest and highest weight, their infinite sum is neither lowest nor highest weight (since the corresponding weights respectively decrease/increase with $n$).
}
Moreover, the tensor product in the relation \eqref{twistFF} hides the need of a suitable completion of the corresponding vector space in the left-hand-side, since the corresponding change of basis relating the two modules involves an infinite linear combination of the
generators.\footnote{See the last paragraph of Section 2 in \cite{Basile:2018dzi} for more comments and details on this subtlety (in higher dimensions).} 
Note that the norm of an infinite linear combination of the generators may diverge, even though each summand has a finite norm, hence the completion of the tensor product of two unitary modules (here, the singleton times its conjugate) may result in a non-unitary module (the adjoint module here).

The left-hand-side stands for the $\mathfrak{so}(2,1)$-module spanned by the endomorphisms of $\cV_{\Delta}$ and can be interpreted concretely in two ways. Firstly, the left-hand-side can be understood as spanned by ``ket-bra'' formed from the elements of the Verma module $\cV_{\Delta}$. Secondly, it can be understood as spanned by elements of the algebra $\mathfrak{gl}[\lambda]$ of endomorphisms of the Verma module $\cV_{\Delta}$. In both case, one obtains the same decomposition \eqref{twistFF} into irreducible $\mathfrak{so}(2,1)$-modules.

\subsection{Abstract definition of the twisted-adjoint action}
\label{sec:twist_general}

Before formulating  the twisted-adjoint action of $\mathfrak{so}(2,1)$ on its universal enveloping algebra, we begin by discussing this construction in the case of any symmetric Lie algebra $\mathfrak{g}$ and its universal enveloping algebra $U(\mathfrak{g})$.

A \textit{symmetric Lie algebra} is a Lie algebra $\mathfrak{g}$ endowed with an involutive automorphism $\tau$. The eigenspaces 
\be
\mathfrak{g}_\pm=\{y\in\mathfrak{g}\,|\,\tau(y)=\pm y\}
\ee 
of $\tau$ provide a decomposition as a semidirect sum $\mathfrak{g}=\mathfrak{g}_+\inplus\mathfrak{g}_-$. In other words, $\mathfrak{g}_+$ is a subalgebra of $\mathfrak{g}$, which will be called the \textit{isotropy subalgebra}, while $\mathfrak{g}_-$ is a $\mathfrak{g}_+$-module,  which will be called the \textit{transvection module}. Equivalently, a symmetric Lie algebra is defined as a Lie algebra which is ${\mathbb Z}_2$-graded in the sense that 
\be
\label{split}
[\mathfrak{g}_+,\mathfrak{g}_+]\subset\mathfrak{g}_+\;,
\qquad
[\mathfrak{g}_+,\mathfrak{g}_-]\subset\mathfrak{g}_-\;,
\qquad
[\mathfrak{g}_-,\mathfrak{g}_-]\subset\mathfrak{g}_+\;.
\ee 
For instance, the isometry algebra $\mathfrak{g}=\mathfrak{so}(d,2)$ of  AdS$_{d+1}$ space  is a symmetric Lie algebra $\mathfrak{g}=\mathfrak{g}_+\inplus\mathfrak{g}_-$ with the Lorentz algebra $\mathfrak{g}_+=\mathfrak{so}(d,1)$ as isotropy subalgebra. Moreover, the complement $\mathfrak{g}_-={\mathbb R}^{d,1}$ is the Lorentz module spanned by the transvections (we avoid to call these  infinitesimal displacements ``translations'' because they do not commute with each other in AdS$_{d+1}$ space). The relations \eqref{split} are then $\mathfrak{so}(d,2)$ commutation relations in the Lorentz basis.

The universal enveloping algebra ${\cal U}(\mathfrak{g})$ of a symmetric Lie algebra $\mathfrak{g}$ is an associative algebra inheriting $\tau$ as a canonical automorphism, which is sometimes called twist. As such it carries two distinct representations of itself. The adjoint representation 
\be
ad:{\cal U}(\mathfrak{g})\to \mathfrak{gl}\big(\,{\cal U}(\mathfrak{g})\,\big):y\mapsto ad_y
\ee
is standard and defined by 
\be
ad_y(a):=y\circ a - a\circ y\;, \qquad \forall a,y\in{\cal U}(\mathfrak{g}) \,,
\ee
where $\circ$ denotes the product in ${\cal U}(\mathfrak{g})$. For a symmetric Lie algebra $\mathfrak{g}$, there is another representation 
\be
\tau ad:{\cal U}(\mathfrak{g})\to \mathfrak{gl}\big(\,{\cal U}(\mathfrak{g})\,\big):y\mapsto \tau ad_y
\ee
of the Lie algebra ${\cal U}(\mathfrak{g})$ on itself, via the definition 
\be
\label{tw_ad}
\tau ad_y(a):=y\circ a-a\circ \tau(y)\,,\qquad \forall a,y\in{\cal U}(\mathfrak{g}) \,,
\ee
which will be called the twisted-adjoint representation.\footnote{The twisted-adjoint representation was originally introduced in the context of higher-spin theory in \cite{Vasiliev:1999ba}. In the mathematical literature it was defined in \cite{Arnaudon}.}

The restrictions of these two representations to the isotropy subalgebra $\mathfrak{g}_+\subset{\cal U}(\mathfrak{g})$ coincide, 
\be\label{filtr}
\tau ad|_{\mathfrak{g}_+}=ad|_{\mathfrak{g}_+}\;.
\ee
However, the restrictions of these two representations to the whole Lie algebra $\mathfrak{g}\subset{\cal U}(\mathfrak{g})$ differ since
\be
\label{tw_adpm}
\tau ad_{y_\pm}(a):=y\circ a\mp a\circ y\,,\qquad \forall a\in{\cal U}(\mathfrak{g}) \,,
\ee
for $y_\pm\in\mathfrak{g}_\pm$.
On the one hand, $ad|_{\mathfrak{g}}$ preserves the canonical filtration of ${\cal U}(\mathfrak{g})$ by the degree in the algebra elements, see Section \bref{sec:adjoint}. On the other hand, $\tau ad|_{\mathfrak{g}}$ preserves the filtration by the degree in the element of the isotropy subalgebra $\mathfrak{g}_+$, cf. \eqref{filtr}, but increases by one the filtration by the degree in the elements of the transvection module $\mathfrak{g}_-$, see Section \bref{sec:twisted}. 

Finally, if a (primitive) ideal Ann$V\subset\univ$ of a (irreducible) $\mathfrak g$-module $V$ is preserved by the automorphism $\tau$, then the latter is a well-defined automorphism of the quotient $\univ/$Ann$V\,\cong \mathfrak{gl}(V)$, which therefore carries an adjoint and twisted-adjoint representation of $\mathfrak g$ via the definitions above applied to equivalence classes. In the case of the singleton module Rac$(d,2)$ of the isometry algebra $\mathfrak{g}=\mathfrak{so}(d,2)$ of AdS$_{d+1}$ space, this construction leads to the adjoint and twisted-adjoint module of the bosonic higher-spin algebra \cite{Vasiliev:2003ev,Vasiliev:2004cm,Iazeolla:2008ix}.

\subsection{Twist automorphism and compensator vector }

Let us represent the $\mathfrak{so}(2,1)$ basis elements in the Lorentz basis, $T_A = (P_a, L)$, where $P_a$ are transvections and $L$ is the Lorentz rotation with the commutation relations
$[L,L] = 0$,  $\,[L,P_a] = \epsilon_{ab}P^b$, and $[P_a,P_b] = \epsilon_{ab}L$, 
where $\epsilon_{ab}$ is the $\mathfrak{so}(1,1)$ Levi-Civita symbol ($a,b=0,1$), cf. \eqref{split}. Then, the twist automorphism acts as 
\be
\label{tau}
\tau (L) = L\;, 
\qquad
\tau(P_a) = - P_a\;.
\ee
To implement this transformation within the $\mathfrak{so}(2,1)$ covariant oscillator approach developed in Section \bref{sec:oscillator} we employ the compensator formalism \cite{Stelle:1979aj}.   
The above splitting into rotation and transvections can be done in an $\mathfrak{so}(2,1)$-covariant manner  by introducing a (constant) compensator $\mathfrak{so}(2,1)$-vector $V^A$ subject to the normalisation condition
\be
\label{compensator}
\qquad V_A V^A =1\;.
\ee 
The standard representative can be chosen as $V^A = (0,0,1)$.  
The Lorentz subalgebra $\mathfrak{so}(1,1)\subset \mathfrak{so}(2,1)$ can be defined as the stability algebra of the compensator. Then, any $\mathfrak{so}(1,1)$-tensor $H^{a_1 ... a_s}$ of rank $s$ can be represented as a rank-$s$  $\mathfrak{so}(2,1)$-tensor $F^{A_1A_2 ... A_s}$ subjected to the $V^A$-transversality condition, 
$V_{A_1} F^{A_1A_2 ... A_s} = 0$ (for a detailed  discussion, see e.g. \cite{Bekaert:2005vh}).

Given $\mathfrak{so}(2,1)$ basis elements $T^A$ we  define the {\it covariantized} Lorentz basis elements as  
\be
\label{PL}
\ba{c}
\dps
L = T^BV_B\;,
\qquad 
P^A = T^A - V^A  V_B T^B\;.
\ea
\ee
By construction, the transvections satisfy the transversality condition $P^A V_A = 0$, so that the standard choice of the compensator reproduces transvections $P^A = (P^a,0)$ in the Lorentz basis. 

The compensator defines a covariant splitting of the auxiliary variables $Z^A$  along  parallel and transversal directions,  
\be
\label{parper1}
Z^A = \Ypar^A+\Yper^A\;, 
\ee
where 
\be
\ba{c}
\Ypar^A = (Z\cdot V)\,V^A\;,
\qquad
\Yper^A = Z^A\,-\,(Z\cdot V)\,V^A\;.
\ea
\ee
The two sets are transversal to each other, $Z_{||}\cdot Z_{\perp} = 0$. 
Further details as well as our conventions and notation can be found in Appendix \bref{app:isomorphism2}. Here, we list operators used in the sequel,
\be
\label{NZB_pp}
\ba{lll}
\dps
N_{\parallel}  = Z_{||}\cdot \p_{||}\;,
\qquad\qquad
&
Z_{||}^2 = Z_{||} \cdot  Z_{||}\;,
\qquad\qquad
&
\Box_{||} = \p_{||}\cdot \p_{||}\;,
\vspace{-2mm}
\\
\\
\dps
N_{\perp}  = Z_{\perp}\cdot \p_{\perp}\;, 
&
Z_{\perp}^2 = Z_{\perp}\cdot  Z_{\perp}\;, 
&
\Box_{\perp} = \p_{\perp}\cdot \p_{\perp}\;.
\ea
\ee
These are counting and trace operators \eqref{NZB} given in $||$ and $\perp$ independent directions. Each set forms its own $\mathfrak{sl}(2,\mathbb{R})$ algebra, as in \eqref{sl2}.

\subsection{Twisted-adjoint action on the universal enveloping algebra module}
\label{sec:twisted}

Following the general definition \eqref{tw_ad}, the twisted-adjoint action of $\mathfrak{so}(2,1)$ on the universal enveloping algebra $\cU(\mathfrak{so}(2,1))$ is given by
\be
\label{tw_gens}
\mathbb{T}_A F\,\equiv\, \llbracket T_{A},  F \rrbracket_* \;\coloneqq  T_{A}* F  - F* \tau(T_{A})\;,
\qquad\;\;
\forall F\in \univ\;,
\ee
or, taking the twist automorphism \eqref{tau} into account, the twisted-adjoint action can be given in the covariantized Lorentz basis as  
\be
\tL F  =[L, F]_* \qquad\; \text{and}\qquad  \tP_A F= \{P_A,F\}_*\;.
\ee
The double line brackets denote the twisted commutator which as follows from the last line  is either commutator or anti-commutator, depending on $L$ or $P_A$ defined in \eqref{PL}.  Using the star-product expressions \eqref{TAB} and the transverse/parallel variables \eqref{parper1}  we find that 
\be
\label{LLPP}
\ba{c}
\dps
\mathbb{L} =  - \epsilon_{ABC} V^A \,Z_\perp^B \,\p_\perp^C\;,
\vspace{3mm}
\\
\dps
\mathbb{P}^A = Z_\perp^A -(N+2)\, \p_\perp^A  \;,
\\

\ea
\ee 
Here, $N = N_{\parallel} + N_\perp$, cf. \eqref{NZB_pp}. We see that the Lorentz rotation acts homogeneously in $Z_\perp$, while transvections act inhomogeneously since $Z^A_\perp$ and $\p^A_\perp$ respectively increase and decrease the degree by one as discussed in the end of Section 
\bref{sec:twist_general}.         

The covariantized twisted generators satisfy the $\mathfrak{so}(2,1)$ algebra commutation relations   
\be
\label{crt}
\begin{gathered}
[\mathbb{L}, \mathbb{P}^A] =  \epsilon^{ABC}V_C \mathbb{P}_B\;,
\vspace{-4mm}
\\ 
\\
[\mathbb{P}^A, \mathbb{P}^B] = \epsilon^{ABC}V_C \mathbb{L}\;.
\end{gathered}
\ee
The  Casimir operator in the twisted-adjoint representation is given by 
\be
\label{casim}
\ba{c}
\mathbb{C}_2  = \half\mathbb{T}_A\mathbb{T}^A =\half\llbracket T_{A},\llbracket T^{A}, \,\cdot\,\rrbracket_*\rrbracket_* \equiv  \mathbb{P}_A\mathbb{P}^A+\mathbb{L}^2\;,
\\
\\
=-\big[Z_\perp^2-(N+2)(N+3)\big](\Box_\perp-1)+(N_{\parallel}+1)(N_{\parallel}+2)\;,
\ea
\ee
where the central dot stands for an element of the universal enveloping algebra, cf. \eqref{cas_ad}.

\subsection{Towards the twisted-adjoint action on the higher-spin algebra}
\label{sec:towards}

The twisted-adjoint generators \eqref{tw_gens} have a well-defined action on the quotient algebra elements  since 
\be
[\mathbb{T}^A, \hd] = 0\;,
\ee  
where $\hd$ is given in \eqref{DF}. In other words, the quotient is an invariant space. Let us denote the projection of the twisted-adjoint generators on the quotient algebra as   
\be
\label{T_lam}
\mathbb{T}^A \longrightarrow \mathbb{T}^A_\lambda \;,
\ee
where the resulting generators depend linearly\footnote{The twisted Casimir \eqref{casim} reduced on the quotient is quadratic in $\lambda$. Therefore, the twisted generators \eqref{T_lam} on the quotient must be at most linear in $\lambda$. Another indirect argument for this is given in Footnote \bref{deformed}.} 
on $\lambda$. An attempt to build the projection $\mathbb{T}^A_\lambda$  using the general formula \eqref{factoring} fails because the subspace of traceless elements is not $\mathbb{T}^A$-invariant,  $[\mathbb{T}^A, \Box] \not\approx 0$. It means that representatives of the quotient algebra should be chosen in some other way. Doing the explicit projection to find $\mathbb{T}^A_\lambda$ could be technically cumbersome because using the canonical basis in the higher-spin algebra is inadequate for analyzing the twisted-adjoint representation.\footnote{\label{deformed}The twisted-adjoint action on $\hs$ can be explicitly built once the quotient algebra \eqref{quotient} is realized in inner coordinates  using e.g. the deformed oscillators \cite{Vasiliev:1989re}. Then, one can directly see that in this case the twist automorphism  defined with respect to the associative product in the deformed oscillator algebra yields the twisted-adjoint action linear in the parameter $\lambda$. The deformed oscillator approach, interesting in its own, will be considered elsewhere. }
 
Thus, along with the canonical basis described in the Proposition \bref{prop:quotient} one will consider a {\it twisted} basis in the higher-spin algebra which is  more suitable when discussing the twisted-adjoint $\mathfrak{so}(2,1)$ action both on the universal enveloping algebra and its quotients. 
 
For our purpose, let $\tilde{\cS}_3$ be a completion of the commutative algebra ${\cS}_3$ of polynomials in the variables $Z$, see \eqref{FZ}. More precisely, the commutative algebra $\tilde{\cS}_3$ is spanned by polynomials in $Z_\parallel$ but one allows for power series in $Z_\perp$.  This completion is important because we propose to replace the trace condition $\Box H(Z) = 0$ for the canonical basis by the modified trace condition $\left(\Box_\perp - 1\right) H(Z) = 0$ for the twisted basis. However, the equation $\left(\Box_\perp - 1\right) H(Z) = 0$ does not possess polynomial solutions apart from the trivial one. Strictly speaking, one should introduce an analogous completion of the universal enveloping algebra $\univ$ such that it is isomorphic to $\tilde{\cS}_3$ as a linear space, but we will keep this subtlety implicit in the sequel. 
\begin{prop}
\label{prop:quotient_H}
As a linear space, the quotient algebra \eqref{quotient} is isomorphic to the infinite-dimensional subspace $\tilde{\cH} \subset \tilde{\cS}_3$ singled out by the constraint $\left(\Box_\perp - 1\right) H(Z) = 0$. This subspace is graded by the homogeneity degree in the variables $Z_{\parallel}^A$,
\be
\label{HS_space_H}
\tilde{\cH} = \bigoplus_{n=0}^\infty \cW_n\;, 
\qquad\quad
\cW_n = \left\{H\in\cS_3\; :\;\, \left(\Box_\perp - 1\right) H = 0\;, \;\; (N_{\parallel}-n)H = 0\right\}\;.
\ee  
\end{prop}

\noindent The detailed proof is given in  Appendix \bref{app:isomorphism2}. Below we just outline the general heuristic idea behind  the Proposition. 

\begin{figure}[H]
\hspace{2cm}\includegraphics[width=0.7\linewidth]{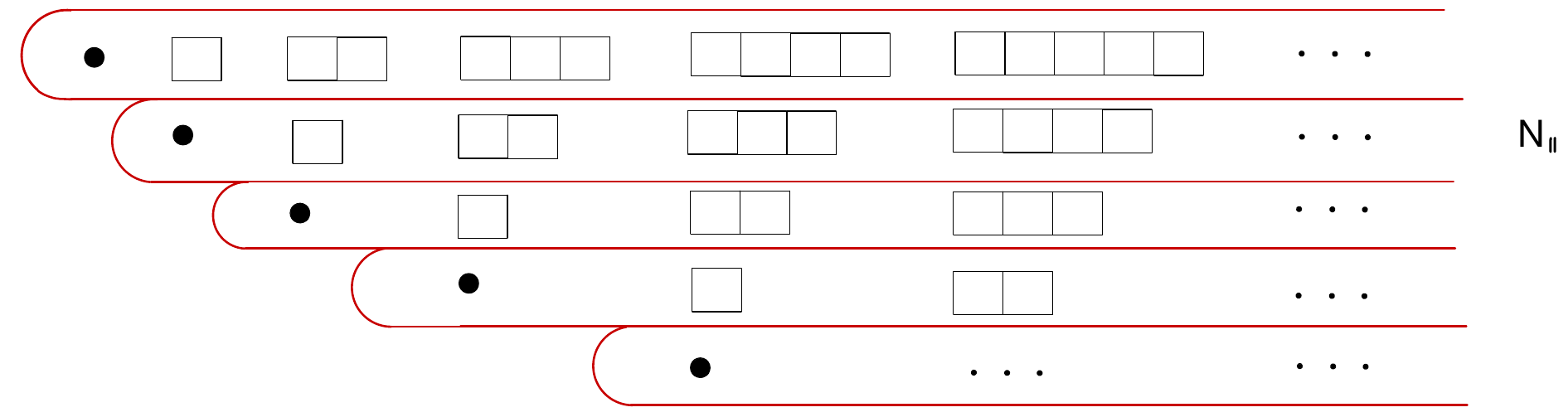}
\caption{The $\mathfrak{so}(1,1)$-covariant basis of $\hs$ in the twisted form. Each Young diagram of length $k$ is an $\mathfrak{so}(1,1)$ rank-$k$ irreducible  module $d_k$. Subspaces $\cW_n$ are shown in red contours. The counting operator $N_\parallel$ has a fixed eigenvalue on all elements inside a given contour.}
\label{fig_univ_red0}
\end{figure}

In the canonical basis, the subspace $\cH\subset\cS_3$ is singled out by  the relation $\Box H(Z)  =0$ which is the three-dimensional
d'Alembert equation. The solution space is known to be given in terms of the irreducible finite-dimensional $\mathfrak{so}(2,1)$-modules $\cD_n$ of all ranks ($n = 0,1,2,...$) or, equivalently, in terms of ``spherical harmonics'': $\cH = \cD_0\oplus \cD_1 \oplus \cD_2 \oplus \ldots\, $, see  \eqref{HS_space}.
Now, the branching rule $\cD_n \cong  d_0 \oplus d_1 \oplus d_2 \oplus \ldots \oplus d_n$ of each irreducible $\mathfrak{so}(2,1)$-modules as a sum of $\mathfrak{so}(1,1)$-modules $d_k$ (of dimensions $\dim d_0=1$ or $\dim d_k=2$ for $k>0$) spanned by rank-$k$ traceless Lorentz tensors, leads to
\be\label{cano}
\cH \equiv \cD_0 \oplus \cD_1 \oplus \cD_2 \oplus \ldots= \infty \cdot d_0\oplus \infty \cdot d_1 \oplus \infty\cdot  d_2 \oplus \ldots\;,
\ee 
where the infinite coefficients in the right-hand-side mean that each $\mathfrak{so}(1,1)$-module $d_k$ enters in a (countably) infinite number of copies.

In the twisted basis, the subspace $\tilde{\cH}\subset\tilde{\cS}_3$ is singled out by the constraint  $(\Box_\perp-1)H(Z) = 0$ which is the two-dimensional Klein-Gordon equation and its general solution has the form of an infinite  expansion in powers of the radius with coefficients being ``spherical harmonics'' on the circle, i.e. Fourier modes, which can be written in covariant form.\footnote{In the context of the continuous-spin field theory, such a modified trace condition was previously considered in \cite{Bekaert:2005in,Alkalaev:2017hvj,Bekaert:2017xin,Alkalaev:2018bqe} and a functional class in oscillator variables analogous to the completion $\tilde\cS_3$ was previously discussed in \cite{Alkalaev:2017hvj}.} In our case, the analogue of the radius is $Z_\perp$ and its square $Z_\perp^2$ is the trace creation operator (in the $\perp$ direction). Since $H(Z) = H(Z_\perp, Z_\parallel)$,  the dependence of the parallel coordinate $Z_\parallel$ is not constrained by the  Klein-Gordon equation so that the homogeneity degree in $Z_\parallel$ simply defines an extra index running from zero to infinity. 
In other words, the space of solutions to the constraint $(\Box_\perp-1)H(Z) = 0$ can be represented as (see Fig. \bref{fig_univ_red0}) 
\be
\label{twis}
\tilde{\cH} \equiv \cW_0 \oplus \cW_1 \oplus \cW_2 \oplus \ldots \;=\; \infty \cdot (d_0\oplus d_1 \oplus d_2 \oplus \ldots)\;.
\ee 
Here, again, the infinite coefficient on the right-hand-side means that each $\mathfrak{so}(1,1)$-module $d_k$ has infinite multiplicity. As a linear space, each infinite-dimensional indecomposable $\mathfrak{so}(2,1)$-module $\cW_k$ decomposes into the infinite sequence $\cW_k \cong d_0\oplus d_1 \oplus d_2 \oplus ... $ of $\mathfrak{so}(1,1)$-module for any $k$.  Now, comparing \eqref{cano} and \eqref{twis}, we arrive at the one-to-one correspondence between the canonical and twisted bases realizing the quotient space.

\begin{figure}[H]
\hspace{2cm}\includegraphics[width=0.7\linewidth]{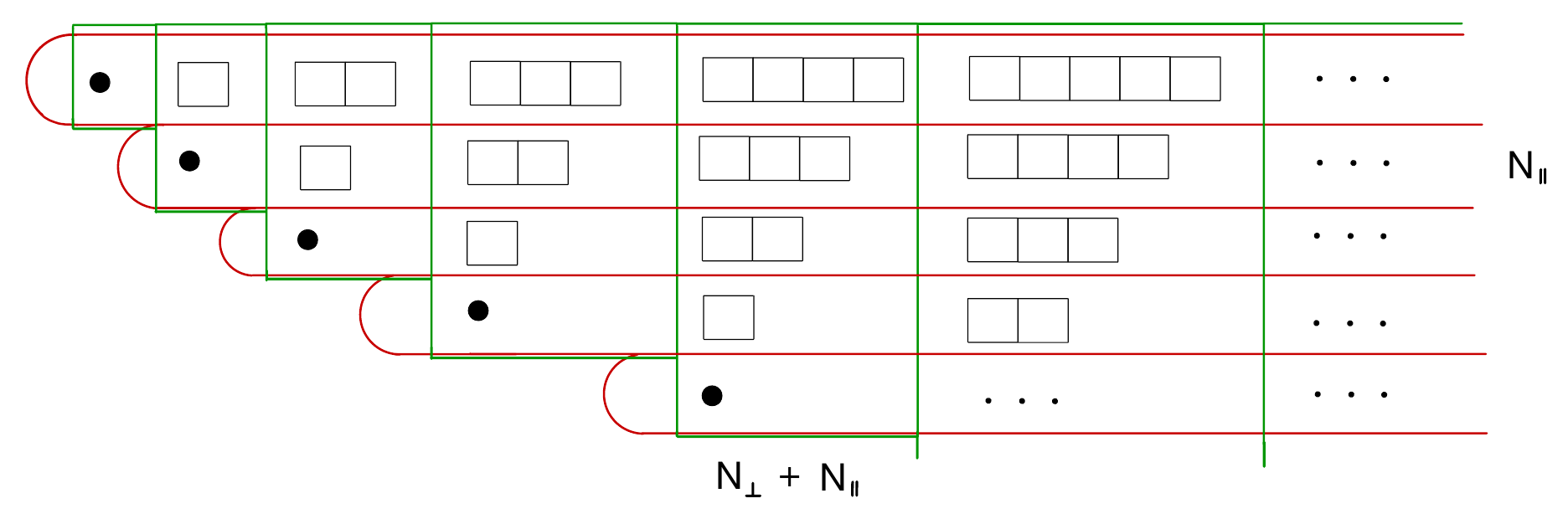}
\caption{The $\mathfrak{so}(1,1)$-covariant basis of $\hs$ in the canonical and twisted forms. Each Young diagram of length $k$ stands for a rank-$k$ traceless $\mathfrak{so}(1,1)$-tensor module $d_k$. Subspaces $\cW_n$ are shown in red horizontal contours. The counting operator $N_{\parallel}$ has a fixed eigenvalue on all elements inside a given red contour. The $\mathfrak{so}(2,1)$-modules  $\cD_n$ are shown in green vertical contours. The counting operator $N=N_\perp+N_{\parallel}$ has a fixed eigenvalue on all elements inside a given green contour.}
\label{fig_slice}
\end{figure}

The twisted-adjoint generators $\mathbb{T}^A_\lambda$ have a well-defined action in each subspace $\cW_n$ thereby endowing them with an $\mathfrak{so}(2,1)$-module structure. In the next section, we will  show that the $\mathfrak{so}(2,1)$-modules $\cW_n$ in \eqref{HS_space_H} are isomorphic to Verma modules $\cV_{\Delta_n}$ with conformal weights  depending linearly on $\lambda$,  
\be
\label{Weyl_Verma}
\cW_n \cong \cV_{\Delta_n}
\qquad
\text{with the weight}
\qquad
\Delta_n = n-\lambda+1\;, 
\qquad
n = 0,1,2,...\;.
\ee
In the higher-spin literature, the modules $\cW_n$ are known as Weyl modules (see e.g. \cite{Bekaert:2005vh}), cf. Section \bref{sec:unfolded}.

\section{Verma modules as twisted-like representations}
\label{sec:verma}

We introduce a class of representations of $\mathfrak{so}(2,1)$ referred to as {\it twisted-like} representations. Such representations  generalize the twisted-adjoint representation of the Lie algebra $\mathfrak{so}(2,1)$ on its universal enveloping algebra $\univ$. These are always infinite-dimensional modules decomposed into an infinite collection of Verma modules with running conformal weights.

\subsection{Twisted-like representations}
 
 The twisted-adjoint generators $\mathbb{L}$ and $\mathbb{P}^A $  belong to the general class of $\mathfrak{so}(2,1)$ generators \eqref{crt} such that transvections are realized  inhomogeneously in $Z_\perp$-variables,   
\be
\label{P_conf}
\mathbb{L} =  - \epsilon_{ABC} V^C \,Z_\perp^A \p_\perp^B\;,
\qquad\qquad 
\mathbb{P}^A = Z_\perp^A -\left[ N_\perp +{\bf \Delta}(N_{\parallel}) \right]\p_\perp^A\;,
\ee
see Appendix \bref{app:twisted} for details.  The commutation relations \eqref{crt} are satisfied for any function ${\bf \Delta}(N_{\parallel})$ which is a power series in $N_{\parallel}$. It will be referred further as the {\it conformal-weight function}.

Note that the above realization of $\mathfrak{so}(2,1)$ on the linear space $\cS_3$ resembles the way the conformal algebra  acts on primary fields of some CFT. In our case, the $\mathfrak{so}(2,1)$ conformal algebra with basis elements $(P,K,D)$ can be realized, on functions $\phi(t)$ of the only ``spacetime'' variable $t\in \mathbb{R}$, as inhomogeneous differential operators  
\be
P = \frac{d}{dt}\;,
\qquad
K = t^2 \frac{d}{dt}+2t \Delta\;,
\qquad 
D = -t \frac{d}{dt} -  \Delta\;,
\ee
where  $\Delta$ is a conformal dimension (see also Appendix \bref{app:verma}). Introducing the counting operator $N=\dps t\frac{d}{d t}$ and defining $L_{\pm} = P\pm K$, $L_0 = -D$ we find  the isomorphic $\mathfrak{sl}(2, \mathbb{R})$ algebra basis elements 
\be
\label{o21sl2}
L_{\pm} = \frac{d}{dt} \pm t(N+ 2\Delta-1)\;,
\qquad 
L_0 = N+\Delta\;,
\ee
which are schematically the same as the twisted-like generators \eqref{P_conf}, i.e. homogeneous $L_0$ and inhomogeneous $L_{\pm}$.

The important property of the twisted-like representations is that the generators \eqref{P_conf} weakly commute with the operators $N_{||}$ and $\Box_{\perp}-1$\,:
\be
\label{constra}
[\mathbb{L}, N_{||}] = 0\;,
\quad
[\mathbb{P}^A, N_{||}] = 0\;,
\quad
[\mathbb{L}, \Box_{\perp}-1]  = 0 \;,
\quad
[\mathbb{P}^A, \Box_{\perp}-1]  =2\,\p_\perp^A (\Box_\perp - 1)\;.  
\ee
Moreover, these two operators trivially commute with each other,
\be
[N_{||},\Box_{\perp}-1]=0\;.
\ee     
Following the Proposition \bref{prop:quotient_H},  the linear space of the higher-spin algebra $\hs$ can be realized as the subspace $\tilde{\cH} \subset \tilde{\cS}_3$ singled out by the modified constraint $(\Box_{\perp}-1)H(Z)=0$ and further sliced according to the counting operator eigenvalues $N_{\parallel}$. From \eqref{constra}, it follows that the twisted-like generators have a well-defined action on $\tilde{\cH}$, thereby endowing it with an $\mathfrak{so}(2,1)$-module structure.

Let $\cH_{{\bf \Delta}}$ be the vector space $\tilde{\cH}$ considered as an $\mathfrak{so}(2,1)$-module specified by the conformal-weight function $\cwf$. The $\mathfrak{so}(2,1)$-module $\cH_{{\bf \Delta}}$ is sliced according to the eigenvalues $n$ of the operator $N_{\parallel}$, where each submodule is isomorphic to the Verma module $\cV_{\Delta_n}$ with the conformal weight $\Delta_n \equiv \Delta(n)$. Thus, $\cH_{{\bf \Delta}}$ is always an infinite direct sum of Verma modules with weights defined by the particular  conformal-weight function,
\be
\label{tA}
\cH_{{\bf \Delta}} = \bigoplus_{n=0}^\infty \cV_{\Delta_n}\;.
\ee
The Casimir operator eigenvalue on the Verma module $\cV_{\Delta_n}$ is given by  
\be
[\mathbb{C}_2]_n = \Delta_n\left(\Delta_n-1\right)\,,
\ee
see \eqref{casimir_n} and \eqref{two_deltas}.

\subsection{Special values and factorization}
\label{sec:special}

By using \eqref{parper1}, any $F\in \cS_3$ can be represented as $F(Z) = F(Z_\perp, Z_{||})$
and, therefore, one can introduce filtrations with respect to degrees in transversal and parallel variables which are  $\perp$ degree and $||$ degree, respectively. The linear space of  $\cS_3$ is then bi-graded as follows
\be
\label{filtration}
\cS_3 = \bigoplus_{n,m=0}^\infty \cU_{n,m}\;,
\ee
where $\cU_{n,m}$ are subspaces of fixed degrees,
\be
\cU_{n,m} = \{\,\forall\, F\in \cS_3\;:\;N_{\parallel} F = n\, F\;, \; N_{\perp} F = m \,F\, \}\;.
\ee

Let us consider transvections $\mathbb{P}^A$ defined by an arbitrary conformal-weight function ${\bf \Delta}(N_{\parallel})$, cf \eqref{P_conf}, which act on the vector space \eqref{filtration} isomorphic to the universal enveloping algebra $\univ$. First, from \eqref{constra} we have $[\mathbb{P}^A, N_{\parallel}] = 0$, and, therefore, $\tP_A$ keeps the $||$ degree invariant. Second, the transvections are inhomogeneous operators that can be split into two parts as 
\be
\label{PWW}
\mathbb{P}^A = \mathbb{W}^A_+ + \mathbb{W}^A_-\;,
\ee
where 
\be
\label{WW}
\mathbb{W}^A_+ = Z_\perp^A\;,
\qquad
\mathbb{W}^A_- = -\left[ N_\perp +{\bf \Delta}(N_{\parallel}) \right]\p_\perp^A \;.
\ee
The operators  $W^A_{\pm}$ respectively increase/decrease the $\perp$ degree, 
\be
\label{W12}
\begin{gathered}
\mathbb{W}^A_{+}\;:\;\;\;  \cU_{n,m} \to \cU_{n,m+1}\;,\\
\;\;\mathbb{W}^A_{-}\;:\;\;\;  \cU_{n,m}\to \cU_{n,m-1}\;,
\end{gathered}
\ee
though their sum  \eqref{PWW} does not have this property, mixing up terms of different $\perp$ degrees. The Lorentz rotation does not change any of the two degrees, 
\be
\label{L_rot}
\mathbb{L}:\; \cU_{n,m} \to \cU_{n,m}\;,
\ee
and, therefore, any subspace $\cU_{n,m}$ is $\mathfrak{so}(1,1)$-invariant.

Since the $||$ degree remains intact under the action of the twisted-like generators we conclude that the eigenvalue $n$ of $N_{\parallel}$ labels different $\mathfrak{so}(2,1)$-modules. For the sake of simplicity, let us keep $n$ implicitly and denote $\cU_{n,m} \equiv \cU_{m}$ implying  that we have chosen a particular $\mathfrak{so}(2,1)$-module with fixed $n$. Thereafter, the number $m$ will be referred to as {\it level}. From \eqref{L_rot} it follows that each level is Lorentz-invariant (i.e. $\mathfrak{so}(1,1)$-invariant), while transvections shift levels back and forth. 

Now, let us define the following filtration by the subspaces $\cQ_{l}$ of the $\perp$ degree not greater than $l\in \mathbb{N}_0$\,: 
\be
\cQ_{l} = \bigoplus_{m=l}^\infty \,\cU_{m}\;\;,
\qquad
\cQ_{l} \subset \cQ_{l-1} \subset ... \subset \cQ_{1}\subset \cQ_{0}\;.
\ee

In general, by virtue of the property \eqref{W12}  the whole space $\cQ_{0}$ is invariant with respect to the action of the transvection 
\be
\mathbb{P}^A\cQ_{0} \subset \cQ_{0}\;.
\ee
However, for special values of the prefactor in the decreasing operator $\mathbb{W}_-^A$ defined in \eqref{WW}, a subspace $\cQ_{l}\subset \cQ_{0}$ may be an invariant subspace for some $l$,
\be
\mathbb{P}^A\cQ_{l}\subset \cQ_{l} \;.
\ee
Note that the operator $\mathbb{W}_+^A$  acts in $\cQ_0$ by raising the grading number \eqref{W12}, hence subspaces $\cQ_l$ are $\mathbb{W}_+^A-$invariant at any $l\in \mathbb{N}_0$. In contrast, the operator $\mathbb{W}_-^A$ acts by lowering the grading number \eqref{W12} and its natural invariant subspaces are the complements to $\cQ_l \subset \cQ_0$. Then, considering their sum  we find out that there are $\tP^A-$invariant subspaces provided that  the prefactor in $\mathbb{W}_-^A$ vanishes for particular eigenvalues of $N_{\perp}$ and of the conformal-weight function ${\bf \Delta}(n)$,
\be
\label{singular}
\mathbb{W}_-^A \cQ_l =0\;: \qquad  l-1+{\bf \Delta}(n)\approx 0\;,
\ee 
where the weak equality symbol $\approx$ means that it holds when acting on polynomials $F(Z_{\perp}, Z_{||})$ with fixed $||$ degree $n$. As the prefactor is linear in $N_\perp$, there are only two eigenspaces of $\mathbb{W}_-^A$ with eigenvalue zero on $\cQ_0$: 

\begin{itemize}
\item[$-$] the subspace $\cU_{0}\subset \cQ_0$ (these are constants in $Z_\perp$ on which $\p_\perp$ acts by zero);

\item[$-$] the subspace $\cU_{l}\subset \cQ_0$ (where the prefactor equals zero \eqref{singular}).  

\end{itemize}

\noindent As a consequence, the quotient  
\be
\label{Dl}
D_{l-1}  = \cQ_{0}/\cQ_{l}
\ee
is a (finite-dimensional) $\mathfrak{so}(2,1)$-module. This is essentially the same mechanism  of quotienting with respect to invariant subspaces as for  Verma modules and singular submodules, as  reviewed in Appendix \bref{app:verma}. 

In general, the $\mathfrak{so}(2,1)$-module $D_{l}$ is reducible because there remains the modified trace constraint to be imposed. Following the discussion below \eqref{trace_schema}, we conclude that the subspace singled out by the modified trace constraint in the quotient $D_l$ is the irreducible finite-dimensional $\mathfrak{so}(2,1)$ module $\cD_l$ \eqref{finite-dim} of dimension $2l+1$.  

By way of illustration, let us consider e.g. the conformal-weight function 
\be
\label{example}
{\bf \Delta}(N_{\parallel}) = N_{\parallel}-\lambda+1\;,
\qquad
\mbox{for} \,\,\lambda \in \mathbb{N}_0\;.
\ee 
Then, for modules  $\cQ_{0}$ labeled by $n$  the equation \eqref{singular} has zeros at the levels $m=\lambda-n$. It follows that the quotient spaces arises in subspaces  with $n = 0,1,2, ...,\lambda-1$.\footnote{Formally, one should also consider $n = \lambda$, but in this case the zero is achieved by acting with $\p_\perp$ on constants  as discussed below \eqref{singular}.}  For instance, fix $\lambda=1$ so that there is just one zero in the module $\cQ_0$ with $n=0$. The resulting quotient $D_0$ is the trivial $\mathfrak{so}(2,1)$-module $\cD_0$ because the zero arises at  level 1, cf. \eqref{finite-dim}.

\begin{figure}[H]
\hspace{2cm}\includegraphics[width=0.7\linewidth]{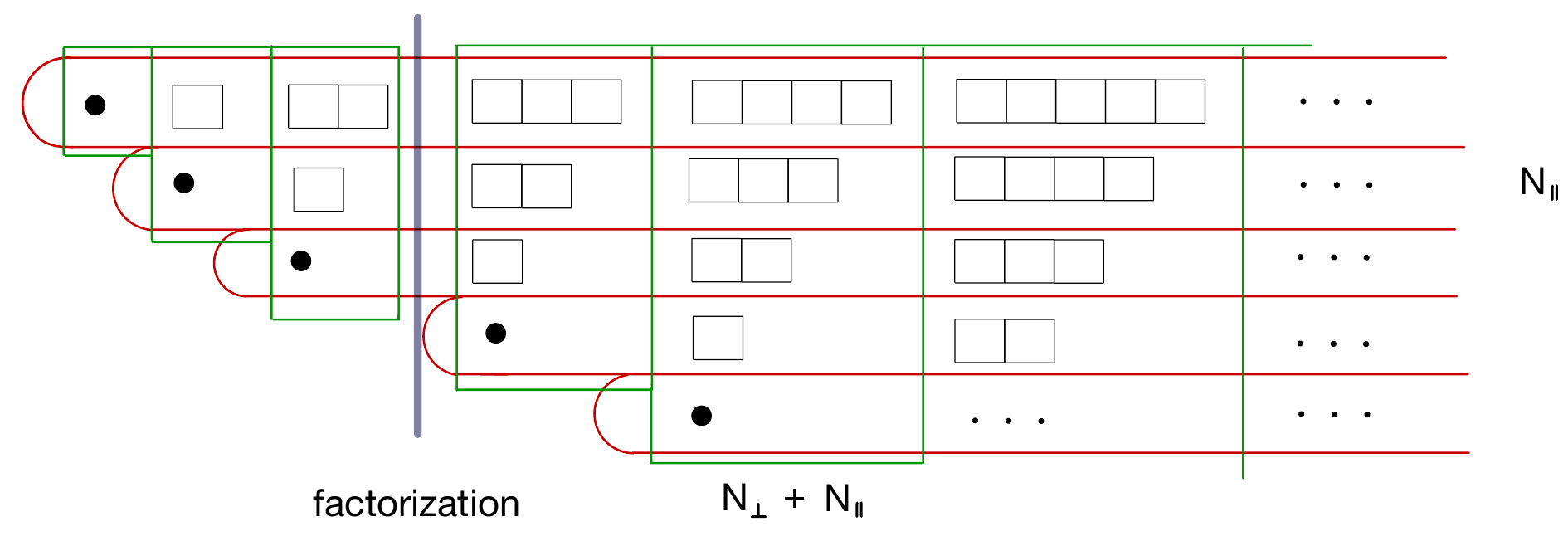}
\caption{The subspace $\cH_{{\bf \Delta}}$ singled out by the modified trace condition is sliced into Verma modules  $\cV_{\Delta_n}$ with $n=0,1,2, ...$, see \eqref{tA}. Green and red contours show $\cD_l$ and $\cV_{\Delta_n}$ modules, respectively. The vertical violet line visualizes  the factorization for the conformal-weight function \eqref{example}.}
\label{fig_factor}
\end{figure}

Further imposing the modified trace constraint $\Box_{\perp}-1 \approx 0$ we arrive at the module $\cH_{{\bf \Delta}}$ \eqref{tA} specified by the conformal-weight function \eqref{example}. The factorization yields the finite-dimensional module 
\be
\label{Q1}
\cD_0 \oplus \cD_1 \oplus \cD_2 \oplus ... \oplus \cD_{\lambda-1}\;,
\ee
which is shown on Fig. \bref{fig_factor} on the left-hand side. On the other hand, the right-hand side is the infinite-dimensional module singled out from the subspace $\cS_\lambda$ by the modified trace constraint,
\be
\label{Q2}
\cD_{\lambda}\oplus \cD_{\lambda+1}\oplus \cD_{\lambda+2}\oplus ... \;.
\ee 
This can be directly seen by counting elements in the basis of $\mathfrak{so}(1,1)$, both in the ideal \eqref{Q2} and in the resulting quotient \eqref{Q1}, see Appendix \bref{app:isomorphism2}.  

\subsection{Twisted-adjoint action on the higher-spin algebra}

As discussed in Section \bref{sec:towards} we do not explicitly build the quotient $\hs$ in terms of inner coordinates. Instead, starting from the universal enveloping algebra $\univ$ as the twisted-like $\mathfrak{so}(2,1)$-module defined by some $\cwf$ and using the factorization properties of $\hs$ at integer $\lambda$, reviewed in Section \bref{sec:extra}, we can fix uniquely how the twisted-adjoint action $\mathbb{T}^A$ on $\univ$ can be projected onto $\hs$ to obtain $\mathbb{T}^A_\lambda$.

In order to build the twisted-adjoint module $\hs$ of $\mathfrak{so}(2,1)$ with generators acting as $\mathbb{T}^A_\lambda$, we proceed as follows. Let us consider a twisted-like action of the $\mathfrak{so}(2,1)$ algebra, i.e. the Lorentz rotation and transvection generators are realized by  \eqref{P_conf} with the conformal-weight function $\cwf$ satisfying the following  conditions:

\begin{enumerate}[label=(\alph*)]

\item The function $\cwf$ is linear in $\lambda\,$; 

\item The prefactor $l-1+\cwf$ has zeros at $\lambda\in \mathbb{Z}$ in line with the factorization \eqref{integer_f}.  

\end{enumerate}

The  condition (a) follows from the fact that the equivalence relation \eqref{DF} is quadratic in $\lambda$ and, therefore, the $\mathfrak{so}(2,1)$ generators acting on the representatives are linear in $\lambda$ (the Lorentz rotation are intact, the transvections are modified). The   condition (b) claims that the twisted-module structure on the equivalence classes must conform the existence of additional ideals in $\hs$ at integer $\lambda$ thereby guaranteeing the isomorphism \eqref{integer_f}. In fact, this last condition uniquely fixes the conformal-weight function  to be  linear also in $N_{\parallel}$. The resulting $\cwf$ is exactly given by  \eqref{example}.  As an example we can examine the  $\lambda=1$ case. Then the quotient
$\mathfrak{gl}[1]/{\cal J}_1\cong{\mathbb R}$ agrees with \eqref{integer_f} that exactly corresponds to $D_0$ from \eqref{Dl}. At general integer $\lambda \in \mathbb{N}$ both the quotient and the  subspace are given by \eqref{Q1} and \eqref{Q2} in accordance with \eqref{gl(N)} and \eqref{ideal}.\footnote{On the other hand, this factorization can be shown by analyzing two types of the twist-invariant ideals in the universal enveloping algebra $\univ$, one of which yields the higher-spin algebra $\hs$ and the other one guarantees the further factorization \eqref{Q1} at integer $\lambda$  \cite{Alkalaev:2014qpa}.}

Finally, we obtain  that the solution is uniquely given by  
\be
\label{TAL}
\mathbb{T}^A_{_{\hspace{-0.5mm}\lambda}}\;:
\qquad 
\mathbb{L}_\lambda =  - \epsilon_{ABC} V^C \,Z_\perp^A \p_\perp^B\;,
\qquad
\mathbb{P}^A_{_{\hspace{-0.5mm}\lambda}} = Z_\perp^A -\left( N_\perp +N_{\parallel}-\lambda+1 \right)\p_\perp^A\;.
\ee
The respective Casimir operator eigenvalue evaluated on the constraints \eqref{HS_space_H} can be read off from \eqref{casimir_n} and takes values
\be
\label{cas_eigen}
\left[\mathbb{C}_2\right]_n = (n-\lambda)(n-\lambda+1)\;, 
\qquad
n = 0,1,2,...\;,
\ee
cf. \eqref{mass_squared}, consistently with conformal weights given by \eqref{dim}.

\section{$\ads2$  massive scalars and twisted-adjoint equations}
\label{sec:unfolded}

Let $\cM_2$ be a two-dimensional manifold with local coordinates $x^m$ ($m=0,1$). To describe gravitational fields in two dimensions,  we  introduce an $\mathfrak{so}(2,1)$-valued connection one-form
\be
W = W^A \, T_A\;,
\qquad
\text{where}
\qquad
W^A  =  dx^m W^A_m\;,  
\ee
and $T^A$  are $\mathfrak{so}(2,1)$ basis elements ($A = 0,1,2$). Using the compensator vector \eqref{compensator} and the covariantized basis \eqref{PL} the connection can be split into the frame (zweibein) field $e$ and the Lorentz spin  connection $\omega$ as the sum $W=e+\omega$ where
\be
e^A =  W^A   - V^A \omega\;,
\qquad
\omega = V_A W^A\;,
\ee  
such that the frame field is a covariantized $\mathfrak{so}(1,1)$-vector field, $V_A\, e^A = 0$. Note that in higher dimensions the Lorentz rotations are given by antisymmetric matrices, but in two dimensions it can be dualized to give a vector. Fixing the compensator as $V^A = (0,0,1)$ we arrive at the standard zweibein $e_m^a$ and spin connection $\omega_m$. The associated curvature two-form is given by 
\be
\label{curv}
R^A = d W^A + \half \epsilon^{ABC} W_B \wedge W_C\;,
\ee
where $d = dx^m \p_m $ is the de Rham differential, and $\wedge$ is the exterior product. 
The metric is defined in the standard fashion as $g_{mn} = \eta_{AB}e_m^A e_n^B$. Then, all derivative objects (like curvatures, covariant derivatives, etc) can be introduced in the standard way.

The $\ads2$ spacetime is described locally by the connection one-form $W$ satisfying the zero-curvature condition 
\be
\label{zero_curv}
R^A(W) = 0\;,
\ee
provided the zweibein $2\times 2$ matrix $e_m^a$ is invertible. Note that the above three equations are equivalent to the single Jackiw-Teitelboim equation $R - \Lambda =0$, where $R$ and $\Lambda$ are the standard scalar curvature and the cosmological constant \cite{Teitelboim:1983ux,Jackiw:1984je,Fukuyama:1985gg}.

\subsection{Covariant derivatives}

Let $\Omega_p(\cM_2)\otimes \tilde{\cS}_3$ be the space of differential $p$-form fields  on the manifold  $\cM_2$ taking values in the totally-symmetric representations of $\mathfrak{so}(2,1)$ with arbitrary ranks in this fibre. These can be packed into generating functions in the auxiliary commuting variables $Z^A$. For arbitrary $U(x|Z) \in \Omega_p(\cM_2)\otimes \tilde\cS_3$ we have a decomposition  
\be
U(x|Z) = \sum_{n=0}^\infty U_{A_1 ... A_n}(x)\, Z^{A_1} ...\, Z^{A_1}\;,
\ee 
where the expansion coefficients are implicitly $p$-forms and explicitly $\mathfrak{so}(2,1)$-tensors. 

The adjoint derivative reads
\be
\label{nabla}
\nabla  = d + W^A \cT_A\;, 
\ee
where $\cT_A$ are basis elements of $\mathfrak{so}(2,1)$ in the adjoint representation \eqref{adj_T}. 
 
We can also introduce a class of covariant derivatives that can be called twisted-like derivatives  as 
\be
\label{nabla_t}
\tilde \nabla  = d + W^A \mathbb{T}_A\;, 
\ee
where $\mathbb{T}_A = (\mathbb{P}_A,\mathbb{L})$ are basis elements of $\mathfrak{so}(2,1)$ in the twisted-like representation \eqref{P_conf}, defined by particular conformal-weight function.  

Both derivatives being squared yield  the curvature 2-from 
\be
\label{nabla2}
\nabla \nabla  = R^A \cT_A\;,
\qquad\;\;
\tilde \nabla \tilde \nabla  = R^A\mathbb{T}_A\;.
\ee  

\subsection{Gauge field equations}

Let $\Omega_1(\cM_2)\otimes \cH$ be the subspace of differential one-forms $\Omega = \Omega(x|Z)$ singled out by the trace constraint $\Box\,\Omega = 0$.
Following the Proposition \bref{prop:quotient}, the solution is given by a collection of $1$-forms taking values in finite-dimensional irreducible representations $\cD_s$ of the algebra $\mathfrak{so}(2,1)$,
\be
\Omega(x|Z) = \sum_{s=0}^\infty \Omega_{A_1 ... A_s}(x) Z^{A_1} ... Z^{A_s}\;,
\ee
where each expansion coefficient is a one-form with totally-symmetric and traceless indices,
\be
\Omega_m{}^{A_1 ... A_s} = \Omega_m{}^{(A_1 ... A_s)}\;,
\qquad
\eta_{A_1A_2}\Omega_m{}^{A_1A_2A_3 ... A_s} = 0\;.
\ee
They are the gauge fields in the adjoint representation arising through the gauging the higher-spin algebra $\hs$. 

The equations of motion are written using the adjoint derivative \eqref{nabla} as the covariant constancy condition 
\be
\label{eom_gauge}
\nabla \Omega \equiv (d+ W^A \cT_A)\, \Omega = 0\;,
\ee  
where the connection one-form describes the $\ads2$ spacetime \eqref{zero_curv}. Therefore, they describe the linearized dynamics in the gauge sector. These equations can be obtained as linearization of the full equations of motion following from the BF action with the gauge algebra $\hs$ \cite{Alkalaev:2013fsa,Alkalaev:2014qpa}. There are no local physical degrees of freedom propagated by the gauge connections because the action is topological while the resulting equations of motion \eqref{eom_gauge} are covariant constancy conditions in finite-dimensional representations (see e.g. discussion in \cite{Vasiliev:1994gr}).

\subsection{Massive scalar field equations}

Let $\Omega_0(\cM_2)\otimes \tilde{\cH}$ be the space of differential $0$-forms $C = C(x|Z)$ singled out by the modified trace constraint 
\be
(\Box_\perp -1)\,C = 0\;,
\ee
which may be even further restricted by fixing the $\parallel$ degree,
\be
(N_{\parallel} - n)\, C = 0\;,
\qquad
n = 0,1,2, ...\;. 
\ee 
The two conditions above can be solved in Lorentz components as a collection of infinite-dimensional spaces labelled by the eigenvalue $n$,
\be
\label{weyl_dec}
\Omega_0(\cM_2)\otimes \tilde{\cH} = \bigoplus_{n=0}^\infty \cW_n\;,
\qquad
\text{where}
\qquad
\cW_n  = \{ \stackrel{(n)}{C}_{a_1 ... a_k}(x)\;|\; k=0,1,2,...\}\;,
\ee
where each element  is a totally-symmetric and traceless $\mathfrak{so}(1,1)$-tensor field, 
\be
\stackrel{(n)}{C}_{a_1 ... a_k} = \stackrel{(n)}{C}_{(a_1 ... a_k)}\;,
\qquad
\eta^{a_1a_2}\stackrel{(n)}{C}_{a_1 ... a_k} = 0\;.
\ee
Following the Proposition \bref{prop:quotient_H} such a collection of fields form the twisted-adjoint representation of the higher-spin algebra $\hs$. 

The equations of motion can be written using the  twisted-adjoint derivative \eqref{nabla_t} as the covariant constancy condition 
\be
\label{cov_const}
\tilde\nabla_{_{\hspace{-0.5mm}\lambda}}C \equiv \left(d + W_A \mathbb{T}^A_{_{\hspace{-0.5mm}\lambda}}\right) C = 0 \;,  
\ee
where again the connection one-forms $W^A$ subjected to the zero-curvature condition  \eqref{zero_curv} describe the $\ads2$ spacetime,  and $\mathbb{T}^A_{_{\hspace{-0.5mm}\lambda}}$ are basis elements of $\mathfrak{so}(2,1)$ in the twisted-adjoint representation \eqref{TAL}. 

Decomposing $C\in \Omega_0(\cM_2)\otimes \tilde{\cH}$ according to \eqref{weyl_dec} we can project the original equation \eqref{cov_const} onto subspaces $\cW_n$ with $n=0,1,2,...$ The resulting equations in each irreducible sector are known as the unfolded equations for scalar fields, while subspaces $\cW_n$ are the Weyl modules \cite{Vasiliev:1995sv,Shaynkman:2000ts}. From the analysis of the previous section  it follows that the Casimir operator eigenvalues on each subspace $\cW_n$ is given by \eqref{cas_eigen}. Then, the system \eqref{cov_const} describes physically an infinite  tower of on-shell $\ads2$ scalar fields $\Phi_n \equiv \stackrel{(n)}{C}(x)$ given by the lowest component in each Weyl module \eqref{weyl_dec}, 
\be
\label{ads_scalar}
\left(\nabla^2+M_n^2\right)\Phi_n = 0\;, 
\qquad 
\text{with masses}
\qquad
M^2_n  =  \frac{(n-\lambda)(n-\lambda+1)}{R^{2}_{_{\hspace{-0.5mm}\rm AdS}}}\;,
\ee
where $\nabla^2$ is the d'Alembert operator on the $\ads2$ spacetime of curvature radius $R_{_{\hspace{-0.5mm}\rm AdS}}$, and $n=0,1,2,...$\,.

\section{Comments on candidate interactions}\label{sec:candidateints}
\label{sec:con}

In higher-dimensional HS gravity theory, already at linear level, both the adjoint and the twisted-adjoint equations form a coupled system of equations that describes propagating degrees of freedom.\footnote{In the literature, this is known as the central on-mass-shell theorem that explains how the two types of equations, for $0$-form and $1$-form HS fields, are glued to each other (for reviews, see e.g. \cite{Bekaert:2005vh}). The resulting physical degrees of freedom correspond to propagating massless HS fields and a massive scalar field which altogether form a HS multiplet.} In two dimensions, the only propagating degrees of freedom are in the matter sector so the linear system decouples into two independent sectors: the gauge sector (topological) and the matter sector (dynamical) described respectively by the adjoint and twisted-adjoint equations.

The massive scalar equations \eqref{ads_scalar} define the quadratic part of the Lagrangian \eqref{lagrange}. It is an interesting and important  problem to build a full interacting theory controlled by the higher-spin algebra $\hs$. Such a theory inevitably involves  dynamical matter interacting through the topological gauge sector described by the $\hs$ BF action \cite{Alkalaev:2013fsa,Alkalaev:2014qpa}. Integrating out the topological degrees of freedom, would leave only infinitely many scalars of the matter sector, governed by a higher-spin invariant Lagrangian of the schematic\footnote{Strictly speaking, integrating out fields with local physical degrees of freedom generically leads to nonlocalities. Although the gauge fields which are integrated out are topological in the present case, it is not obvious that the corresponding effective Lagrangian would be of the simple local form \eqref{lagrange}. We ignored this subtlety in the above argument for the sake of simplicity.} form \eqref{lagrange}. At the level of equations of motion, a non-linear theory could be built by employing the rich algebraic structures underlying higher-spin interactions known in higher dimensions \cite{Vasiliev:1990vu,Vasiliev:1992av,Vasiliev:2003ev,Alkalaev:2014nsa,Vasiliev:2018zer,Sharapov:2019vyd}.

\vspace{7mm}
\noindent \textbf{Acknowledgements.} We are grateful to N.~Boulanger, A.~Campoleoni, M. Grigoriev, D. Grumiller, C.~Iazeolla, A.~Latyshev, K.~Mkrt\-chyan, C.~Peng, D.~Ponomarev, A.~Sharapov, K.~Susuki, E.~Skvortsov, A.~Yan, J.~Yoon, for useful discussions. 

X.B. is grateful to LPI Moscow for hospitality in summer 2018 where this project was initiated. K.A. is grateful to AEI Potsdam and IDP University of Tours for their hospitality in winter 2019 where a part of this work was done. We acknowledge the APCTP Pohang for  
hospitality in October 2019 where those results were announced. The work of K.A. was supported by the RFBR grant No 18-02-01024  and by the Foundation for the Advancement of Theoretical Physics and Mathematics “BASIS”.

\appendix

\section{$\mathfrak{sl}(2,{\mathbb R})$ representation theory}
\label{app:verma}

Here we recap some basic facts from the $\mathfrak{sl}(2,{\mathbb R})$ representation theory (see e.g. \cite{Kitaev:2017hnr,Mazorchuk}). 

\subsection{$\mathfrak{sl}(2,{\mathbb R})$ Verma modules }

Let $E,F,H$ be basis elements of the Lie algebra $\mathfrak{sl}(2,{\mathbb R})$ satisfying the Chevalley-Serre relations\footnote{In this subsection, one will follow the standard conventions from mathematical textbooks, e.g. \cite{Mazorchuk}.}
\be
[H,E] = +2E\;,
\qquad
[H,F] = -2F\;,
\qquad
[E,F] = H\;.
\ee 

The Verma modules of $\mathfrak{sl}(2,{\mathbb R})$ can be explicitly constructed as follows. Let $V_\alpha$ be an infinite-dimensional vector space with basis elements $v_0, v_1,v_2,...$\,. The action of the Lie algebra $\mathfrak{sl}(2,{\mathbb R})$ on the representation space $V_{\alpha}$ is a morphism $\pi_\alpha\hspace{-1mm} :\,\mathfrak{sl}(2,{\mathbb R})\to\mathfrak{gl}(V_{\alpha})$ of Lie algebras defined by 
\begin{align}
\pi_\alpha(F)v_n & = v_{n+1}\label{1st}\;,\\
\pi_\alpha(H)v_n & = (\alpha - 2n)v_{n}\;,\label{2nd}\\
\pi_\alpha(E)v_n & = n(\alpha-n+1)v_{n-1}\;,\label{3rd}
\end{align}
where $n \in \mathbb{N}_0$. Here, $v_0$ is the highest-weight vector with $\alpha\in \mathbb{C}$ being its weight. All  $v_n$ are linearly independent by construction. 

From \eqref{3rd} we see that the prefactor is quadratic in $n$, implying that there are two vectors which could be annihilated by $\pi_\alpha(E)$: at $n=0$ this is the highest-weight vector $v_0$ and at $n=\alpha+1$ there is possibly a singular vector $v_n$. It follows that if $\alpha=N-1$ is a non-negative integer (i.e. $N\in{\mathbb N}$), then the vectors $v_{N},v_{N+1},...$ span an invariant infinite-dimensional subspace isomorphic to the Verma module $V_{-N-1}$ of negative highest-weight. The quotient space 
\be
\label{D}
D_N = V_{N-1}\,/\,V_{-N-1}\;,
\ee    
is the irreducible finite-dimensional $\mathfrak{sl}(2,{\mathbb R})$-module of dimension $N$. The negativity of the weight $-N-1<0$ guarantees that there are no further invariant subspaces. 

\subsection{$\mathfrak{so}(2,1)$ Verma modules}

Let us now rewrite the above construction for the conformal algebra $\mathfrak{so}(2,1)\cong \mathfrak{sl}(2,{\mathbb R})$. The  algebra $\mathfrak{so}(2,1)$ in the conformal basis $\{P,D,K\}$ is given by the commutation relations 
\be
[D,P] =P\;,
\qquad
[D,K] = -K\;,
\qquad
[P,K] = 2D\;.
\ee
It can be realized by the vector fields
\be
P = \frac{d}{d x}\;,
\qquad
K = -x^2 \frac{d}{d x}
\;,
\qquad 
D = -x\,\frac{d}{d x}
\;,
\ee
acting on functions $\phi(x)$ on the real line ($x\in \mathbb{R}$) of conformal weight  $\Delta$. The isomorphism $\mathfrak{sl}(2,{\mathbb R})\cong \mathfrak{so}(2,1)$ is achieved by the identification
\be
E\mapsto K\;,
\qquad
F\mapsto P\;,
\qquad
H\mapsto -2D\;. 
\ee 
Accordingly, the Verma module $V_\alpha$ of $\mathfrak{sl}(2,{\mathbb R})$ with highest weight $\alpha$ will be identified with the
Verma module $\cV_{\Delta}$ of $\mathfrak{so}(2,1)$ with lowest conformal weight $\Delta=-\alpha/2$, and the finite-dimensional $\mathfrak{sl}(2,{\mathbb R})$-module \eqref{D} will be identified with the finite-dimensional $\mathfrak{so}(2,1)$-module \eqref{finite-dim}. 
The vector $v_0$ is then identified with a primary field, while the vectors $v_{n>0}$ are the descendants.  

\subsection{$\mathfrak{so}(2,1)$ lowest-weight unitary irreducible representations}

To discuss the unitary irreducible representations of $\mathfrak{so}(2,1)$, it is useful to take inspiration from the textbook example of $\mathfrak{so}(3)$. For this reason, one will make use of quantum-mechanics notations.\footnote{We closely follow the conventions and notations from the review \cite{Kitaev:2017hnr}, except for the sign of the quadratic Casimir operator and for the symbol of the conformal weight.} The generators will be denoted $\{L_{-1},L_0,L_{+1}\}$ and the commutation relations $[L_m,L_n]=(m-n)L_{m+n}$ are consistent with the identification
\be
L_{-1}\mapsto -K\;,
\qquad
L_0\mapsto -D\;,
\qquad
L_{+1}\mapsto P\;,
\ee 
which corresponds to their realization as vector fields on the real line $L_n=x^{1-n}\frac{d}{dx}$.
One considers an orthonormal basis where the element $|m\rangle$ is an eigenvector of $L_0$ with eigenvalue $-m$.  

The Casimir operator is equal to
\be
{\cal C}_2\big(\mathfrak{so}(1,2)\big)=L_0^2-\frac12(L_{+1}L_{-1}+L_{-1}L_{+1})\,.
\ee
The eigenvalue of the Casimir operator on a \textit{unitary} irreducible module must be a \textit{real} number.
On a Verma module $\cV_{\Delta}$ with lowest conformal weight $\Delta$,
the value of the Casimir operator is equal to $\Delta(\Delta-1)$. 
Using the commutation relations and the expression for the Casimir operator, one finds that
\begin{align}
\label{irrme1}
\bra{m}L_{+1}L_{-1}\ket{m}&=m(m-1)+\Delta(1-\Delta)=(m-\Delta)(m-1+\Delta),\\[3pt]
\label{irrme2}
\bra{m}L_{-1}L_{+1}\ket{m}&=m(m+1)+\Delta(1-\Delta)=(m+\Delta)(m+1-\Delta).
\end{align}
Unitarity and the relation $L_n^\dagger=L_{-n}$ impose that the right-hand-sides of the above equations must be non-negative for all values of $m$ that appear in the representation.
For a lowest-weight Verma module $\cV_{\Delta}$ the corresponding eigenvalues of $L_0$ are $-m=\Delta,\,\Delta+1,\,\Delta+2,\,...$ and one can check that the necessary and sufficient condition for the right-hand-side of \eqref{irrme1} and \eqref{irrme2} to be non-negative is that the lowest conformal weight $\Delta$ is non-negative.
Up to some arbitrary phase factors which can be absorbed in the definition of the basis vectors, one finds that the orthonormal basis for $\Delta>0$ is given by
\begin{equation}
\label{L_irrep}
\begin{aligned}
L_{-1}\ket{m}&=-\sqrt{(m-\Delta)(m-1+\Delta)}\,\ket{m-1},\\[3pt]
L_0\ket{m}&=-m\,\ket{m},\\[3pt]
L_{+1}\ket{m}&=-\sqrt{(m+\Delta)(m+1-\Delta)}\,\ket{m+1},
\end{aligned}
\end{equation}
with $m=\Delta,\,\Delta+1,\,\Delta+2,\,...$

\section{Basic star-product expressions}
\label{app:star}

We are interested in products like $T_{AB}*F$ (or $F*T_{AB}$) as they define the left (or right) action of the algebra $\mathfrak{so}(2,1)$. Below we explicitly calculate such expressions, both in original doublet variables $Y_\alpha^A$ and in singlet variables $Z^A$. In the doublet variables we get the left and right actions  
\be
\label{TF}
T_{AB} * F(Y) = \frac{1}{8}\left[4\epsilon^{\alpha\beta} Y_{\alpha}{}_A Y_{\beta}{}_B  \,\textcolor{Red}{-}\, 2L_{AB} + \epsilon^{\alpha\beta}\frac{\p^2}{\p Y^\alpha{}^{A} \p Y^\beta{}^B }\right]F(Y)\,,
\ee
\be
\label{FT}
F(Y)*T_{AB} = \frac{1}{8}\left[4\epsilon^{\alpha\beta} Y_{\alpha}{}_A Y_{\beta}{}_B  \,\textcolor{ForestGreen}{+}\, 2 L_{AB} +\epsilon^{\alpha\beta} \frac{\p^2}{\p Y^\alpha{}^{A} \p Y^\beta{}^B }\right]F(Y)\,,
\ee
where 
\be
\label{LAB}
L_{AB} = Y_\alpha{}_A \frac{\p}{\p Y_\alpha^B} - Y_\alpha{}_B \frac{\p}{\p Y_\alpha^A}\;. 
\ee

As an example of passing from $Y_\alpha^A$ to $Z^A$ variables, for the last term in \eqref{TF} and  \eqref{FT} we have 
\be
\epsilon^{\alpha\beta}\frac{\p^2}{\p Y^\alpha{}^{A} \p Y^\beta{}^B} F(Y) = 2\epsilon_{ABC} \left(2+Z^K\frac{\p}{\p Z^K}\right)\frac{\p }{\p Z_C} F(Z)\;,
\ee
or, for the second term, 
\be
\label{lab}
L_{AB} = Z_A \frac{\p}{\p Z^B} -Z_B \frac{\p}{\p Z^A} \;.
\ee

Introducing the Hodge dualized $\mathfrak{so}(2,1)$ basis elements 
\be
\label{TA}
T_A  = \epsilon_{ABC}T^{BC}\;,
\ee
the relations \eqref{TF} and \eqref{FT} can be equivalently rewritten in terms of the singlet variables as
\be
\ba{c}
\label{TAB}
\dps
T^{A} * F(Z) = \frac{1}{2}\left[Z^A -\left(2+Z^B \frac{\p}{\p Z^B}\right)\p^A \textcolor{Red}{-}\,\epsilon^{ABC}Z_B\frac{\p}{\p Z^C} \right]F(Z) \;,
\\
\\
\dps
F(Z)*T^{A} = \frac{1}{2}\left[Z^A -\left(2+Z^B \frac{\p}{\p Z^B}\right)\p^A \textcolor{ForestGreen}{+}\,\epsilon^{ABC}Z_B\frac{\p}{\p Z^C} \right]F(Z) \;.
\ea
\ee

\section{Canonical basis}
\label{app:isomorphism}

\paragraph{Proof of Proposition \bref{prop:quotient}.}  Let $G_{n}\in \cS_3$ be an element of maximal degree $n$ in the variables $Z^A$. It can be decomposed as 
\be
\label{GFG}
G_n = F_n + H_{n-1}\;,
\ee
where $F_n$ is {\it homogeneous} of degree $n$, i.e. $N  F_n = n  F_n$,  while  $H_{n-1}$ is of maximal degree $n-1$.  The expansion coefficients in the homogeneous part,
\be
\label{FZn}
F_n = F_{A_1 ... A_n} Z^{A_1} ...\, Z^{A_n}\;,
\ee  
form a rank-$n$ totally-symmetric $\mathfrak{gl}(3, \mathbb{R})$-tensor, cf. \eqref{FZ}. Further decomposing $F_n$ into $\mathfrak{so}(2,1)\subset \mathfrak{gl}(3, \mathbb{R})$ irreducible components, we find that
\be
\label{trace_dec_GFn}
F_n = T_n + Z^2 F_{n-2}\;,
\ee
where 
\be
\Box T_n = 0\;,
\qquad
N T_n = n T_n\;,
\qquad
N F_{n-2} = (n-2) F_{n-2}\;.
\ee
Here,  $\Box$, $Z^2$, and $N$ are operators introduced in \eqref{NZB}. Now, using the explicit form of the equivalence operator $\hd$ \eqref{DF} we obtain 
\be
G_n = T_n - 16\, \hd F_{n-2} + G_{n-1}\;.
\ee
Here, the term $G_{n-1}$ is of maximal degree $n-1$ so one can apply to $G_{n-1}$ the same logic that was applied to $G_{n}$ in \eqref{GFG}, etc. Eventually, for any element  $G\in \cS_3$ of arbitrary degree we get the trace decomposition 
\be
\label{trace_dec_GF}
G = T - 16 \,\hd F\;,
\qquad
\Box T = 0\;,
\ee 
with some  $F\in\cS_3$. Now, recalling that the gauge equivalence operator reads $\hd F = (C_2 - \mu_\lambda) *F$ and successively applying the relation \eqref{trace_dec_GF} we can arrive at the star-product decomposition \eqref{factoring}.

\begin{figure}[H]
  \centering
    \includegraphics[width=0.8\linewidth]{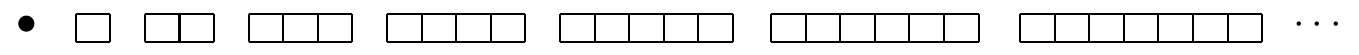}
\caption{The canonical basis in the higher-spin algebra $\hs$. The basis elements are organized into an infinite series of  rank-$n$ traceless totally-symmetric $\mathfrak{so}(2,1)$-tensors ($n=0,1,2, ...$) depicted as one-row Young diagrams of length $n$.    
}
\label{fig_univ_hs}
\end{figure}

Thus, the quotient  \eqref{quotient} defined by the equivalence relation \eqref{equivalence}  is isomorphic to the space of traceless elements as stated in the Proposition \bref{prop:quotient}. Recalling now that the trace and counting operators form an algebra, i.e. $[\Box, N] = 2\,\Box$ from \eqref{sl2}, we conclude that the space $\cal H$ of traceless elements is graded with respect to the eigenvalues $n$ of the counting operator $N$. Thus, we arrive at the decomposition \eqref{HS_space}, cf the plot on Fig. \bref{fig_univ_hs}.

\begin{figure}[H]
  \centering
    \includegraphics[width=0.8\linewidth]{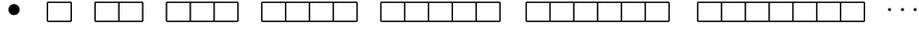}
\caption{The $\mathfrak{gl}(3, \mathbb{R})$-covariant basis in the universal enveloping algebra $\univ$. One-row Young diagrams of length $\ell$ denote here rank-$\ell$ totally-symmetric $\mathfrak{gl}(3, \mathbb{R})$-tensors ($\ell=0,1,2, ...$).
}
\label{fig_univ_tr}
\end{figure}

\begin{figure}[H]
  \centering
    \includegraphics[width=0.8\linewidth]{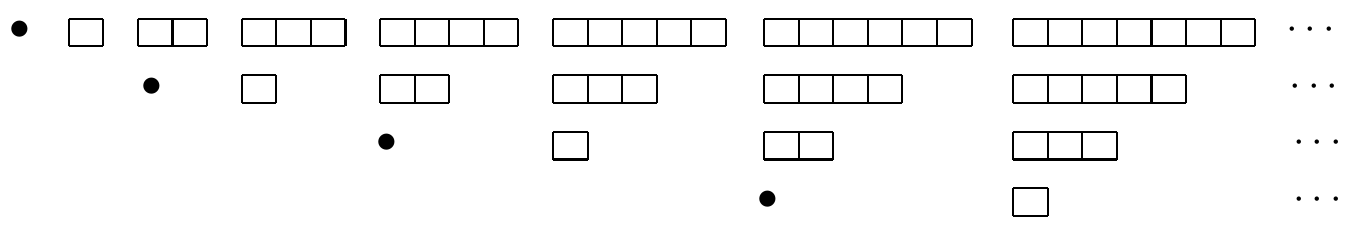}
\caption{The $\mathfrak{so}(2,1)$-covariant basis of the universal enveloping algebra $\univ$. The basis elements are organized into horizontal lines of traceless elements (of ranks running from zero to infinity) for fixed $m$. When combined  into vertical columns (of fixed sum $\ell=2m+n$), the corresponding totally-symmetric traceless $\mathfrak{so}(2,1)$-tensors form a single rank-$\ell$ totally-symmetric $\mathfrak{gl}(3, \mathbb{R})$-tensor on the $\ell$-th position on  Fig. \bref{fig_univ_tr}. Note that Young diagrams in each column indeed differ by two boxes, which corresponds to taking trace of $\mathfrak{gl}(3, \mathbb{R})$-tensors. 
}
\label{fig_univ}
\end{figure}

\paragraph{Proof of Proposition \bref{prop:adj_module}.} First, let us rewrite the decomposition \eqref{FZ} as 
\be
\label{FZA}
\forall F\in \cS_3:\qquad  F = \sum_{\ell\geqslant 0} F_{\ell}\;, \qquad  N F_{\ell} = \ell F_{\ell}\;.
\ee  
Each term $F_\ell = F_{A_1 ... A_\ell} Z^{A_1} ...\, Z^{A_\ell}$ here encodes a totally-symmetric $\mathfrak{gl}(3, \mathbb{R})$-tensor $F_{A_1 ... A_\ell}$ of rank $\ell = 0,1,2, ...$ This is because the linear space $\cS_3$ of the universal enveloping algebra $\univ$ is isomorphic to the symmetric tensor product of $\mathfrak{so}(2,1)$-vectors, hence it is a $\mathfrak{gl}(3, \mathbb{R})$-module. The decomposition \eqref{FZA} can be  depicted as on the plot in Fig. \bref{fig_univ_tr}.  
\noindent Combining the first-trace decomposition \eqref{trace_dec_GFn} and the rank decomposition \eqref{FZA},  we get the general trace decomposition,
\be
\label{general_tr}
\forall F\in \cS_3:\qquad F = \sum_{m,n\geqslant 0} (Z^2)^m\, T_{n}^{(m)} \;,
\qquad
\Box T_{n}^{(m)} = 0\;, \quad N T_{n}^{(m)} = n T_{n}^{(m)}\;, 
\ee 
where the index $n=0,1,2,...$ gives the rank of $T_{n}^{(m)}$ while the extra index $m=0,1,2,...$ labels the tower of such rank-$n$ elements, each tensor $T_{n}^{(m)}$ encoding the information about the $m$-th trace of $F_{n+2m}$ in \eqref{FZA}. Each term in \eqref{general_tr} encodes a rank-$n$ traceless totally-symmetric $\mathfrak{so}(2,1)$-tensor through
\be
\label{Tnm}
T_{n}^{(m)}  = T^{(m)}_{A_1 ... A_n} Z^{A_1} ... Z^{A_n}\;, 
\qquad
\eta^{A_1A_2}T^{(m)}_{A_1 A_2 A_3 ... A_n} = 0\;.
\ee  
Thus, the universal enveloping algebra  $\univ$ can be represented  in the $\mathfrak{so}(2,1)$-covariant basis shown on  Fig. \bref{fig_univ}.

\noindent Following the Proposition \bref{prop:quotient} the linear space of the quotient algebra \eqref{quotient} can be visualized as the first horizontal line on the plot in Fig. \bref{fig_univ} (cf. Fig. \bref{fig_univ_hs}) while all subsequent lines form the ideal. 

Now, let us demonstrate that each copy $\cD^{(m)}_n$ ($m=0,1,2,...$) of the finite-dimensional module $\cD_n$, which appears  in the decomposition \eqref{mult}, indeed carries an adjoint representation of $\mathfrak{so}(2,1)$ spanned by rank-$n$ totally-symmetric traceless $\mathfrak{so}(2,1)$-tensors. To this end, we first note  that $\cD^{(m)}_n$, expressed in terms of the trace expansion \eqref{general_tr}, is spanned by the traceless elements $T^{(m)}_n$. Second, the adjoint action \eqref{adj_T} of $\mathfrak{so}(2,1)$-generators $T_A$ transforms each index as an $\mathfrak{so}(2,1)$-vector and commutes with the counting and trace operators \eqref{NZB}: 
\be
\label{Tsl2}
[\cT^A,\Box] = 0\;, 
\qquad
[\cT^A,Z^2] = 0\;,
\qquad
[\cT^A,N] = 0\;.
\ee
This already shows the vector spaces $\cD^{(m)}_n$ are finite-dimensional $\mathfrak{so}(2,1)$-modules, spanned by rank-$n$ traceless totally-symmetric $\mathfrak{so}(2,1)$-tensors, and are thus isomorphic to $\cD_n$. Let us confirm this by computing the eigenvalues of the Casimir operator of $\mathfrak{so}(2,1)$. In the adjoint representation \eqref{adj}, it reads 
\be
\label{cas_ad}
\cC_2[\mathfrak{so}(2,1)] = \half\cT_A\cT^A  \equiv \half [T_A,[T^A, \,\cdot \,]_*]_* = -Z^2 \Box + N(N+1)\;,
\ee
where the central dot means $\forall F \in \univ$. Then, calculating \eqref{cas_ad} on traceless elements $T^{(m)}_n$ \eqref{Tnm} we find the standard Casimir eigenvalue 
\be
\cC_2[\mathfrak{so}(2,1)]T^{(m)}_n = n(n+1)T^{(m)}_n\;.
\ee

\section{Twisted basis}
\label{app:isomorphism2}

To show the proposition \bref{prop:quotient_H}, we first elaborate the oscillator technique combining  the original variables $Z^A$ and the compensator vector $V^A$ into a single framework that allows  working with Lorentz $\mathfrak{so}(1,1)$-tensors in a manifestly $\mathfrak{so}(2,1)$-covariant manner. 

{\it Parallel/Transverse variables.} Denoting the inner product between two arbitrary $\mathfrak{so}(2,1)$ vectors as $Q\cdot P = \eta_{AB}Q^A P^B$ we can  decompose the original variable $Z^A$ into two parts, parallel and transversal to the compensator vector,    
\be
\label{parper}
Z^A = \Ypar^A+\Yper^A\;, 
\qquad
\text{where}
\qquad
\left\{
  \begin{array}{l}
    \Ypar^A = (Z\cdot V)V^A\;, 
    \vspace{2mm}
    \\ 
    \Yper^A = Z^A-(Z\cdot V)V^A\;. \\
  \end{array}
\right.
\ee
As a consequence, one obtains the transversality relations
\be
\label{trans_cond}
\Yper\cdot V=0\;,
\qquad
\Yper\cdot \Ypar=0\;.
\ee
Introducing  the projector $\Pi_A^B  =\delta_A^B - V^AV_B$ we can define the directional derivatives
\be
\label{proj}
\ba{l}
\dps
\p_A^{\perp} \equiv \frac{\p }{\p \Yper^A} = \Pi_A^B\frac{\p}{\p Z^B}\;,
\qquad
\p_A^{||} \equiv\frac{\p }{\p \Ypar^A} = V_A \left( V\cdot\frac{\p}{\p Z}\right)\;.
\ea
\ee
They satisfy the following relations 
\be
\p_A = \p_A^{\perp}+\p_A^{||}\;,
\qquad
\frac{\p \Yper^B}{\p \Yper^A} = \Pi_A^B\;,
\qquad
\frac{\p \Ypar^B}{\p \Ypar^A} = V_AV^B\;,
\qquad
\frac{\p \Ypar^B}{\p \Yper^A} = 0\;.
\ee
Another relation useful in practice is $Z_{||}^2 \Box_{||} = N_{\parallel}(N_{\parallel}-1)$.

\vspace{3mm}

{\it Tensor analysis.} Any irreducible $\mathfrak{so}(1,1)$-tensor $F_{a_1 ... a_n}$ ($a_i=0,1$) can be represented in an $\mathfrak{so}(2,1)$-covariant way as
\be
\label{Zperp}
F_n(Z_\perp) = F_{A_1... A_n}\, Z^{A_1}_\perp ... Z^{A_n}_\perp\;:
\qquad
\Box_\perp F_n = 0\;, 
\qquad
N_\perp F_n = n F_n\;.  
\ee
By construction, due to \eqref{trans_cond}, the expansion coefficients can be taken to be $V^A$-transverse since any $V^A$-longitudinal piece would lead to a vanishing contribution. Choosing the compensator in the standard form $V^A = (0,0,1)$ one can see that the original rank-$n$ totally-symmetric $\mathfrak{gl}(3, \mathbb{R})$-tensor $F_{A_1... A_n}$ goes to $F_{a_1 ... a_n}$ which is a rank-$n$ totally-symmetric and traceless $\mathfrak{so}(1,1)$-tensor.  On the other hand, any element depending only on  the parallel variables
\be
F_n(Z_\parallel) = F_{A_1... A_n}\, Z^{A_1}_\parallel ... \,Z^{A_n}_\parallel\;
\ee
is a $\mathfrak{so}(1,1)$-scalar. Indeed, introducing the notation 
\be
Z_0 \equiv Z\cdot V
\qquad\text{so that}\qquad 
Z_\parallel^A = V^A Z_0\;,
\ee 
we find that 
\be
\label{Z0}
F_n(Z_\parallel) = F_n(V)\, Z_0^n\;,
\qquad
\text{where}
\qquad
F_n(V) = F_{A_1... A_n}V^{A_1} ... V^{A_n}\;.
\ee
 
Now, we can decompose any element of degree $n$ as 
\be
\label{add}
F_n(Z) = F_{A_1 ... A_n} Z^{A_1} ... Z^{A_n} \equiv \sum_{p+q=n} F_{B_1... B_p| C_1 ... C_q}\, Z^{B_1}_\perp ... Z^{B_p}_\perp\,
Z^{C_1}_\parallel ... Z^{C_q}_\parallel\;. 
\ee
Going from the first equality to the second one, we decompose the expansion coefficients $F_{A_1 ... A_n}$ in $V^A$-transversal and $V^A$-longitudinal parts. The resulting coefficients $F_{B_1... B_p| C_1 ... C_q}$ are $V^A$-transversal tensors with respect to the first group of indices and $V^A$-longitudinal in the second group of indices (i.e. the tensors are proportional to $V_{C_1} ... V_{C_q}$). Using the variable $Z_0$ introduced above, the expansion \label{add} can be represented as
\be
\label{FnZ}
F_n(Z) = \sum_{p+q=n} f^{(q)}_{B_1... B_p}\, Z^{B_1}_\perp ... Z^{B_p}_\perp\, Z_0^q\;,
\ee
where $f^{(q)}_{B_1... B_p}$ satisfy the $V^A$-transversality condition at any $q=0,1,2,...$

\vspace{3mm}

{\it Branching rules.} Imposing the trace condition $\Box F_n = 0$, we see that the expansion coefficients \eqref{FnZ} become related to each other since  the $\mathfrak{so}(2,1)$-trace operator decomposes as  $\Box = \Box_\perp + \Box_\parallel$. Disentangling the resulting system, one finds all irreducible $\mathfrak{so}(1,1)$-tensors of ranks $k = 0,1,2,...,n$ thereby reproducing the branching rule 
\be
\label{brule}
\cD_n = d_0\oplus d_1 \oplus ... \oplus d_n \,,
\ee
where $d_k$ denotes an irreducible $\mathfrak{so}(1,1)$-module spanned by rank-$k$ traceless totally-symmetric Lorentz tensors.
Note that $d_0$ is of dimension one (and carries the trivial representation) while the $\mathfrak{so}(1,1)$-modules $d_k$ are of dimension two for $k>0$. The $\mathfrak{so}(1,1)$-modules $d_k$ are irreducible over the real numbers but become reducible over the complex numbers where they split into the Fourier mode of eigenvalues $\pm k$. In this way, the $\mathfrak{so}(1,1)\subset \mathfrak{so}(2,1)$ branching rule \eqref{brule} is a sum over Fourier modes $k=-n,-n+1,...,-1,0,1,...n-1,n\,$. This is the obvious analogue of the standard $\mathfrak{so}(2)\subset \mathfrak{so}(3)$ branching rule where spherical harmonics are decomposed in terms of eigenvectors of the angular momentum along the vertical axis.

By way of example, let us consider lower-rank elements $F_1$ and $F_2$. For $F_1$, expanding both the coefficient $F_A$ and the variable $Z^A$ in $\perp$ and $\parallel$ parts we get
\be
F_1(Z) = F_{A}Z^A = (F^{\perp}_A+F^{\parallel}_A) (Z^{A}_\perp+Z^{A}_\parallel) = F^{\perp}_A Z^{A}_\perp +F^{\parallel}_A Z^{A}_\parallel \equiv f_A Z^{A}_\perp +f Z_0\;,  
\ee 
cf. \eqref{FnZ}. Fixing the compensator in the standard form we get $F^{\perp}_A \sim F_a$ and $F^{\parallel}_A \sim F$ which are $\mathfrak{so}(1,1)$-tensors of ranks 1 and 0. The analogous procedure for $F_2$ yields  
\be
\label{F2Z}
\ba{c}
F_2(Z) = F_{AB}Z^AZ^B = (F_{AB}^{\perp} + V_{(A} F^{\perp}_{B)}+ V_AV_B F^{\perp})(Z^{A}_\perp+Z^{A}_\parallel)(Z^{B}_\perp+Z^{B}_\parallel)
\\
\\
\dps
 \equiv  (f_{AB} Z^{A}_\perp Z^{B}_\perp) + 2(f_{A} Z^A_\perp) Z_0 + f Z_0^2 \;, 
\ea
\ee
with the expansion coefficients being  $F_{AB}^{\perp} \sim F_{ab}$, $F_{A}^{\perp}\sim F_a$, and $F^\perp \sim F$, which are $\mathfrak{gl}(2, \mathbb{R})$-tensors of respective ranks 0, 1, 2. 

Imposing the trace condition $\Box F_2(Z) = 0$ we find that $\eta^{AB} f_{AB} + f = 0$, i.e. the scalar component arising as the trace is not independent. By appropriate linear  change of basis, the rank-2 tensor can be made traceless. Using the trace relation  obtained above along with the $V^A$-transversality condition we can decompose 
\be
f_{AB} = \stackrel{\circ}{f}_{AB} -\half \Pi_{AB}f\;,
\ee 
where the first term $\stackrel{\circ}{f}_{AB}$ is $\eta^{AB}$-traceless and $V^A$-transversal, and $\Pi_{AB}$ is the projector \eqref{proj}. Then, the expansion \eqref{F2Z} goes to
\be
\label{F2Zfin}
F_2(Z) = (\stackrel{\circ}{F}_{AB}^\perp Z^{A}_\perp Z^{B}_\perp) + 2(F_{A}^\perp Z^A_\perp) Z_0 + F^\perp (Z_0^2 - \half Z_\perp^2) \;. 
\ee 
In other words, the solution to the trace constraint is given by $\mathfrak{so}(1,1)$-components: a rank-2 traceless tensor, a vector, and a scalar. All together they form  an irreducible representation of $\mathfrak{so}(2,1)$ which corresponds to a traceless rank-2 $\mathfrak{so}(2,1)$-tensor $F_{AB}$, thereby demonstrating the $\mathfrak{so}(1,1) \subset \mathfrak{so}(2,1)$ branching rule \eqref{brule}.

\vspace{3mm}

\paragraph{Proof of Proposition \bref{prop:quotient_H}.} Let us introduce two  commuting operators
\be
\label{twist_con}
\Box_\perp - 1\quad \text{and} \quad N_{\parallel}\;:\qquad [N_{\parallel},\Box_\perp-1]=0\;,
\ee  
and consider the subspace $\tilde{\cH}\subset \tilde{\cS}_3$ singled out by the modified trace constraint, 
\be
\label{mtc}
(\Box_\perp - 1) F(Z) = 0\;.
\ee
This is essentially the 
two-dimensional Klein-Gordon equation whose solutions are explicitly known. In what follows we describe the solution space using the technique of $\perp$ and $\parallel$ variables. To this end, let us expand $F(Z)$ in powers of the parallel variable $Z_0$ as in \eqref{FnZ}, 
\be
\label{form}
F(Z) = \sum_{n\geqslant 0} f_{n}(Z_\perp) Z_0^n\;,
\ee
where $f_{n}(Z_\perp)$ are arbitrary power series in $\perp$ variables while the expansion in $Z_0$ can be assumed finite, i.e. $F$ is a polynomial in $Z_0$ but a power series in $Z_\perp$. Such a representation \eqref{form} is convenient because $\perp$ and $\parallel$ variables are naturally separated. Moreover, the solutions $F(Z)$ are graded with respect to the eigenvalues of  the counting operator $N_{\parallel}$ which is now given by 
\be
N_\parallel = Z_0 \frac{\partial}{\partial Z_0}\;.  
\ee

Imposing the modified trace constraint \eqref{mtc} and noticing that $[\Box_\perp, Z_0]=0$, we find solutions of the form  \eqref{form} with expansion coefficients given by \cite{Alkalaev:2017hvj} \be
\label{tr_sol}
f_n(Z_\perp) = \sum_{k\geqslant 0}g_k(Z_\perp^2)f_{n|k}(Z_\perp)\;,
\ee
where the irreducibility conditions 
\be
\label{alpha}
\Box_\perp f_{n|k} = 0\;,
\qquad 
N_{\perp} f_{n|k} = k f_{n|k}\;,
\ee
encode rank-$k$ traceless totally-symmetric $\mathfrak{so}(1,1)$-tensors,
and
\be
\label{trace_schema}
g_k(Z_\perp^2)=\sum_{m= 0}^\infty \frac{4^{-m+1}}{m!(k+\half)_m}\, (Z_\perp^2)^m=4{}_0F_1(k+\tfrac12;\frac{Z_\perp^2}2)
\ee
denote the infinite series of trace creation operators $Z_\perp^2$ (where $(x)_m$ is the Pochhammer symbol) which can be packed into the confluent hypergeometric limit function ${}_0F_1$. Such  collective trace operators act on traceless  elements $f_k(Z_\perp)$, where $k$ is the eigenvalue of $N_\perp$ to produce  elements $g_k(Z_\perp^2) \,f_k(Z_\perp)$ with no definite $\perp$ degree which are however all bounded from below by $k$ with $k=0,1,2,...$\,.        
 
In conclusion, the solution space can be  organized as an infinite number of collections of traceless totally-symmetric $\mathfrak{so}(1,1)$-tensors with ranks running over $k=0,1,2, ...$, encoded in the polynomials $f_{n|k}(Z_\perp)$. Each such collection is labeled by the eigenvalue $n = 0,1,2, ...$  of the counting operator $N_{\parallel}$.

\begin{figure}[H]
\hspace{2cm}\includegraphics[width=0.65\linewidth]{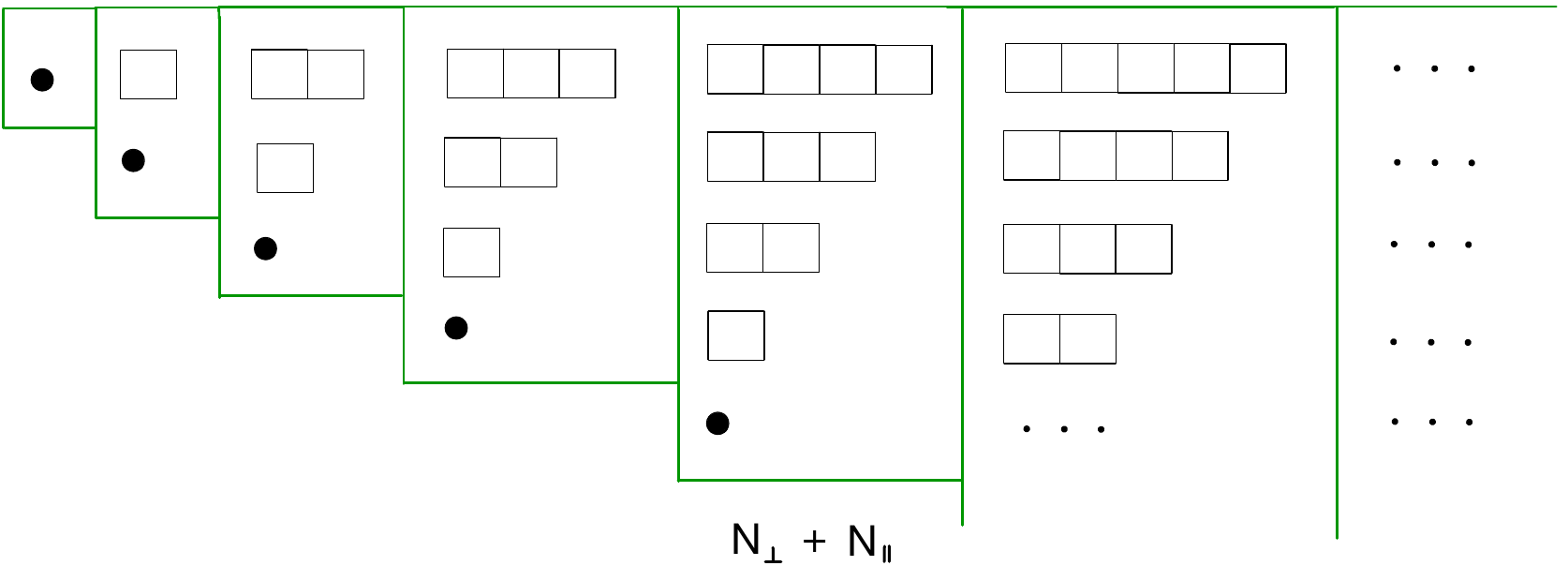}
\caption{The $\mathfrak{so}(1,1)$-covariant basis of $\hs$ in the canonical form. Each Young diagram of length $k$ stands for an irreducible $\mathfrak{so}(1,1)$-module $d_k$. The $\mathfrak{so}(2,1)$-modules  $\cD_n$ are shown in green contours. The counting operator $N=N_\perp+N_{\parallel}$ has a fixed eigenvalue on all elements inside a given contour.}
\label{fig_univ_green}
\end{figure}

\begin{figure}[H]
\hspace{2cm}\includegraphics[width=0.7\linewidth]{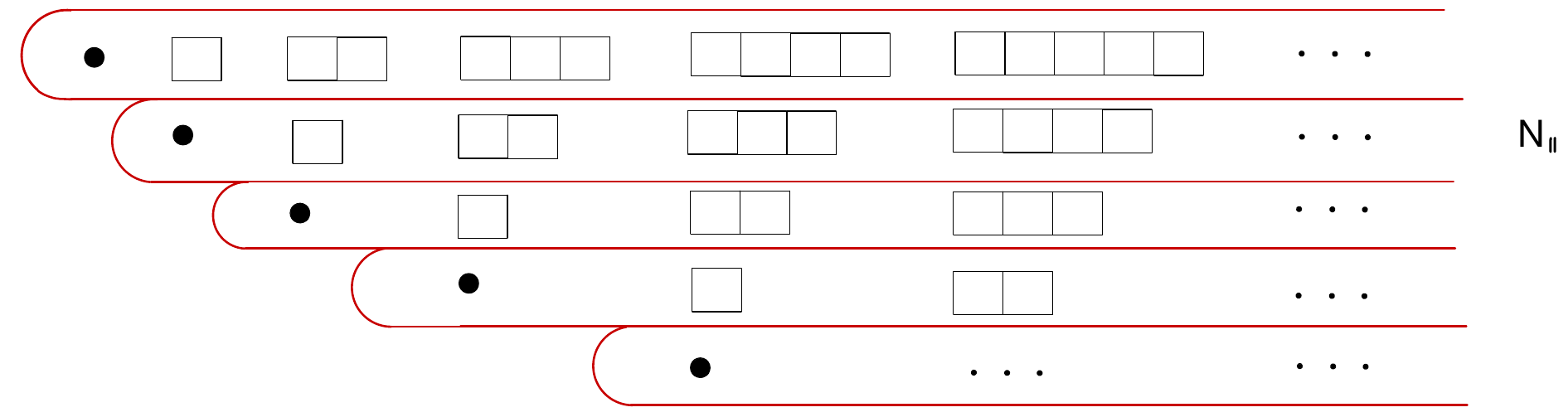}
\caption{The $\mathfrak{so}(1,1)$-covariant basis of $\hs$ in the twisted form. Each Young diagram of length $k$ is a rank-$k$ irreducible  $\mathfrak{so}(1,1)$-module $d_k$. Subspaces $\cW_n$ are shown in red contours. The counting operator $N_{\parallel}$ has a fixed eigenvalue on all elements inside a given contour.}
\label{fig_univ_red}
\end{figure}

{\it Canonical vs twisted basis.}
Let us now recall that, according to the Proposition \bref{prop:quotient},  the linear space of the higher-spin algebra $\hs$ is given in the canonical basis by $\mathfrak{so}(2,1)$-modules $\cD_n$ shown on the plot Fig. \bref{fig_univ_hs}. Since each $\cD_n$ can be decomposed as \eqref{brule} in terms of irreducible $\mathfrak{so}(1,1)$-modules $d_k$, an $\mathfrak{so}(1,1)$-covariant basis of $\hs$ is given by 
\be
\label{brackets}
\ba{c}
\dps
\cH = \bigoplus_{n=0}^\infty \cD_n =  (d_0)\oplus(d_0\oplus d_1)\oplus (d_0\oplus d_1\oplus d_2)\oplus (d_0\oplus d_1\oplus d_2\oplus d_3)\oplus ...\;.
\ea
\ee 
Terms in brackets form $\cD_n$ subspaces, see Fig. \bref{fig_univ_green}. 
\noindent However, the Lorentz modules can be reorganized by introducing an infinite number of subspaces 
\be
\cW_n \cong \bigoplus_{k=0}^\infty d_k\;,
\ee  
where each space $\cW_n$ comprises all the traceless $\mathfrak{so}(1,1)$-tensors $f_{n|k}(Z_\perp)$ which are packed inside a generating function $f_n(Z_\perp)$, cf. \eqref{tr_sol}-\eqref{alpha}. Thus, we obtain a different description of the $\mathfrak{so}(1,1)$-covariant basis of $\hs$ as formulated in the Proposition   \bref{prop:quotient_H},
\be
\tilde{\cH}  = \bigoplus_{n=0}^\infty \cW_n\;,
\ee  
where each subspace $\cW_n$ is the solution to the modified trace and counting constraints \eqref{twist_con}. This is the {\it twisted} basis shown on the plot Fig. \bref{fig_univ_red}.

\noindent The canonical and twisted bases are each $\mathfrak{so}(2,1)$-covariant but with respect to distinct representations. Note that they are related by a triangular linear transformation involving infinite sums. This triangular transformation would involve finite sums if it were done in the space of Lorentz $\mathfrak{so}(1,1)$-components before imposing any $\mathfrak{so}(2,1)$ (standard vs modified) trace condition, as is manifest for instance in Fig. \ref{fig_slice}.

\section{Twisted-like $\mathfrak{so}(2,1)$ representations}
\label{app:twisted}

We assume that the Lorentz generator is uniquely realized by the homogeneous operator 
\be
\label{Lor_g}
\mathbb{L} =  - \epsilon_{ABC} V^C \,Z_\perp^A \p_\perp^B\;,
\ee  
while the transvections 
\be
\label{PAgf}
\mathbb{P}^A = g\,Z_\perp^A +f\, \p_\perp^A
\ee
are defined by some functions $g = g(N_{\parallel}, N_{\perp})$ and $
f = f(N_{\parallel}, N_{\perp})$ 
of the counting operators  \eqref{NZB_pp}. The  coefficient functions are fixed by the requirement that    
\begin{itemize}

\item  the generators $\mathbb{L}$ and $\mathbb{P}_A$ satisfy the $\mathfrak{so}(2,1)$ algebra commutation relations \eqref{crt}. 
\end{itemize}
It turns out that, under the above condition, the resulting generators possess  the following two properties: 

\begin{itemize}

\item[(A)] The generators $\mathbb{L}$ and $\mathbb{P}_A$ preserve two irreducibility conditions: they weakly commute with the weight operator $N_{\parallel}-n$ and the modified trace operator $\Box_{\perp} - h$, where $n\in \mathbb{N}$ and $h = h(N_{\parallel}, N_{\perp})$ is some counting operator (e.g. a constant), i.e.
\be
\label{const}
(N_{\parallel}-n)F = 0\;, \qquad \text{and} \qquad (\Box_{\perp} - h)F =0\;. 
\ee 
Denoting the generators $\mathbb{T}_A = (\mathbb{L},\mathbb{P}_A)$ and the operators $\mathbb{G} = (N_{\parallel}-n,\Box_{\perp} - h)$ the weak commutativity property then reads
\be
[\mathbb{T}_A, \mathbb{G}] \sim \mathbb{G}\;.
\ee 

\item[(B)] The Casimir operator $\mathbb{C}_2 = \mathbb{L}^2+\mathbb{P}_A\mathbb{P}^A$ is diagonalized on the subspace singled out by the irreducibility conditions \eqref{const}. 

\end{itemize}

\noindent Let us introduce the notations
\be
\label{defs}
\begin{gathered}
\,[Z_{\perp}^A, g] = g'\, Z_{\perp}^A\;,
\\
[\p_{\perp}^A, g] = \dot g\, \p_{\perp}^A\;, 
\end{gathered}
\ee
where by dotted and primed functions we denote the results of commuting the counting functions with $\p_\perp$ and $Z_\perp$ respectively.
More explicitly,
\be\label{defss}
g'(N_{\parallel}, N_{\perp})=g(N_{\parallel}, N_{\perp}-1)-g(N_{\parallel}, N_{\perp})\,,\quad
\dot g(N_{\parallel}, N_{\perp})=g(N_{\parallel}, N_{\perp}+1)-g(N_{\parallel}, N_{\perp})
\ee
Now, we consider the three conditions formulated above in turn. 

$\bullet$  The {\it $\mathfrak{so}(2,1)$ commutation relations} are satisfied provided that\footnote{In particular, this condition demonstrates why the coefficient functions are chosen to be independent of the trace  operators in \eqref{NZB_pp}. Otherwise, the first condition  imposes a constraint claiming that a particular combination of  trace operators must act trivially on any element $F(Z)$. Such a constraint singles out a  subspace where the commutation relations of the $\mathfrak{so}(2,1)$ algebra are closed. On the other hand, this is distinct from the generators \eqref{PAgf} which are required to span the $\mathfrak{so}(2,1)$ algebra on any elements of $\univ$.}   
\be
\label{algebra}
g f'  - f \dot g = 1\;.
\ee

$\bullet$ The {\it property (A)} for the counting constraint $N_{\parallel}-n\approx 0$ (the weak equality symbol $\approx$ means equal when the constraints \eqref{const} hold) is satisfied automatically since $N_{\parallel}$ commutes with any $Z^A_\perp$-dependent functions, while $Z_{||}^A$ enter the generators only through $N_{\parallel}$ itself. For the modified trace constraint  $\Box_\perp - h$, we obtain $[\mathbb{L}, \Box_\perp-h] = 0$ and 
\be
\label{trace_cond}
\begin{gathered}
[\mathbb{P}^A, \Box_\perp -h] = -2(g+\dot g + \ddot g)\,\p_{\perp}^A - 2(\dot g +\ddot g) \,Z_{\perp}^A \Box_\perp   - (2 \dot f +\ddot f)\, \p_{\perp}^A \Box_\perp 
\\
\\
 - g h' \,Z_\perp^A - f\dot h\, \p_\perp^A\;.
\end{gathered}
\ee
For particular $f,g$ the right-hand side can be made proportional to $\Box_\perp -h$.

$\bullet$  The {\it property (B)} yields  
\be
\label{Cas_gen}
\mathbb{C}_2 = -\big[Z_\perp^2 - (f^2 + f \dot f)  \big]\left[\Box_\perp -(g^2 + g g') \right] +\mathbb{E}\;,
\ee 
where 
\be
\label{E}
\mathbb{E} = (f^2+ f \dot f)(g^2+g g')+(gf+gf')N_{\perp} +(fg+f\dot g)(N_\perp+2) + N_{\perp}^2\; 
\ee 
purely depends on the counting operators. The first group of terms on the right-hand-side of \eqref{Cas_gen} should weakly vanish. The form \eqref{Cas_gen} of the Casimir operator suggests that the trace constraint must be $\Box_\perp -(g^2 + g g') \approx 0$, while the counting term $\mathbb{E}$ must be  independent of $N_{\perp}$ (otherwise,  $\mathbb{C}_2$ is not diagonal after imposing the trace and weight constraints).

Assuming that the coefficient functions are power series in the counting operators we can represent them in the form  
\be
\label{two_power_series}
\begin{gathered}
g = g_0 +g_1 N_{\perp}+g_2 N_{\perp}^2 + \ldots\;, 
\\
f = f_0 +f_1 N_{\perp}+f_2 N_{\perp}^2 + \ldots\;, 
\end{gathered}
\ee 
where the expansion coefficients $f_i = f_i(N_{\parallel})$ and $g_j = g_j(N_{\parallel})$  depend only on the transverse counting operator $N_{\parallel}$. 

Now, the condition \eqref{algebra} in each order in the variable $N_\perp$ yields an equation on $g_i$ and $f_k$. In other words, there is a single algebraic relation for two functions.  A particular solution fixes one half of the expansion coefficients $f_i$ and $g_j$.  The other half parameterises  equivalent realisations of the $o(2,1)$ generators since such a  freedom can be absorbed by linear changing basis. Thus, one of functions can be chosen arbitrarily while the other is then fixed by means of the relation \eqref{algebra}.

A convenient choice\footnote{The other (opposite) choice would be a constant function $f$.} is a constant function $g$, i.e. 
\be
\label{sol1}
g(N_{\parallel}, N_{\perp}) = \zeta\;, \quad\text{where}\quad \forall\zeta\in \mathbb{C}\;.
\ee
From \eqref{defss}, it follows that all derivative functions are zero:  $g' =0$, $\dot g=0$, etc. Hence, \eqref{algebra} reduces to the relation $f' = \zeta^{-1}$ that can be solved as
\be
\label{sol2}
f = -\zeta^{-1} N_{\perp}+ f_0\;,
\ee
where $f_0 = f_0(N_{\parallel})$ is  arbitrary, as follows from the definition of the prime $'$ in \eqref{defss}. Also, from \eqref{sol2} we directly  find that  $\dot f  = -\zeta^{-1}$, $ f^\prime  = 0$, etc. 

The solution \eqref{sol1}  and the Casimir operator \eqref{Cas_gen} suggest that the trace condition should be given as 
\be
\label{trace2}
\Box_\perp - \zeta^2 \approx 0\;,
\ee 
i.e. $h = \zeta^2$ is an arbitrary constant.\footnote{In this way, we arrive at the modified trace \eqref{trace2} of the same form as in the continuous-spin field theory (see Appendix \bref{app:isomorphism}).} Then, the commutator \eqref{trace_cond} drastically simplifies to yield the required relation 
\be
\label{calc}
[\mathbb{P}^A, \Box_\perp -\zeta^2] = -2g\p_\perp^A -2 \dot f \p_{\perp}^A \Box_\perp  = \frac2{\zeta}\,\p_\perp^A (\Box_\perp - \zeta^2)\approx 0\;,
\ee
which means  that the transvections have a well-defined action on the subspace singled out by the modified trace constraint \eqref{trace2}.

Finally, consider the Casimir operator \eqref{Cas_gen}. Using the trace constraint we see that the first term vanishes. Then, the second term $\mathbb{E}$ \eqref{E}  should give the expected eigenvalue. Indeed, using \eqref{sol1} and \eqref{sol2} we get 
\be
\mathbb{E}= \zeta f_0 \,\big(\zeta f_0 +1\big)\;.
\ee
All in all, we find that the Casimir eigenvalue on each subspace singled out by  the trace constraint $\Box_\perp \approx \zeta^2$ and the counting constraint $N_\parallel\approx n$ parameterized by $n\in \mathbb{N}$ is given by 
\be
\label{casimir_n}
\left[\mathbb{C}_2\right]_n \approx \zeta f_0(n) \,\big(\zeta f_0(n) +1\big)\;.
\ee
Since  the only invariant combination of parameters is $\zeta f_0$, the constant $\zeta$ can be set to unity or absorbed into $f_0$. Therefore, from now on we set $\zeta=1$. Recalling that the $\mathfrak{so}(2,1)$ Casimir eigenvalue is parameterized as $\Delta(\Delta-1)$ and introducing for the conformal weight  the notation ${\bf\Delta}(n)$ we find out that it can be identified as 
\be
\label{two_deltas}
{\bf\Delta}(n) = -f_0(n)\;, 
\ee
where 
\be
\label{f0}
f_0(n) = \alpha_0 + \alpha_1 n +\alpha_2 n^2 + ... \;,
\ee
with arbitrary coefficients $\alpha_i$, $i\in \mathbb{N}$. The dual (shadow) weight is then ${\bf\Delta}(n) = f_0(n)+1$. The resulting expressions for the twisted-like $\mathfrak{so}(2,1)$ generators are given in \eqref{P_conf}.


\begin{thebibliography}{10}

\bibitem{Sachdev:1992fk}
S.~Sachdev and J.~Ye, \emph{{Gapless spin fluid ground state in a random,
  quantum Heisenberg magnet}},
  \href{http://dx.doi.org/10.1103/PhysRevLett.70.3339}{\emph{Phys. Rev. Lett.}
  {\bf 70} (1993) 3339}, [\href{http://arxiv.org/abs/cond-mat/9212030}{{\tt
  cond-mat/9212030}}].

\bibitem{Kitaev}
A.~Kitaev, \emph{{A simple model of quantum holography}}, {\emph{KITP strings
  seminar and Entanglement 2015 program (Feb. 12, April 7, and May 27, 2015)}
  {\bf \url{http://online.kitp.ucsb.edu/online/entangled15/}} }.

\bibitem{Maldacena:2016hyu}
J.~Maldacena and D.~Stanford, \emph{{Remarks on the Sachdev-Ye-Kitaev model}},
  \href{http://dx.doi.org/10.1103/PhysRevD.94.106002}{\emph{Phys. Rev.} {\bf
  D94} (2016) 106002}, [\href{http://arxiv.org/abs/1604.07818}{{\tt
  1604.07818}}].

\bibitem{Sarosi:2017ykf}
G.~Sárosi, \emph{{AdS$_{2}$ holography and the SYK model}},
  \href{http://dx.doi.org/10.22323/1.323.0001}{\emph{PoS} {\bf Modave2017}
  (2018) 001}, [\href{http://arxiv.org/abs/1711.08482}{{\tt 1711.08482}}].

\bibitem{Rosenhaus:2018dtp}
V.~Rosenhaus, \emph{{An introduction to the SYK model}},
  \href{http://arxiv.org/abs/1807.03334}{{\tt 1807.03334}}.

\bibitem{Teitelboim:1983ux}
C.~Teitelboim, \emph{{Gravitation and Hamiltonian Structure in Two Space-Time
  Dimensions}},
  \href{http://dx.doi.org/10.1016/0370-2693(83)90012-6}{\emph{Phys. Lett.} {\bf
  126B} (1983) 41--45}.

\bibitem{Jackiw:1984je}
R.~Jackiw, \emph{{Lower Dimensional Gravity}},
  \href{http://dx.doi.org/10.1016/0550-3213(85)90448-1}{\emph{Nucl. Phys.} {\bf
  B252} (1985) 343--356}.

\bibitem{Fukuyama:1985gg}
T.~Fukuyama and K.~Kamimura, \emph{{Gauge Theory of Two-dimensional Gravity}},
  \href{http://dx.doi.org/10.1016/0370-2693(85)91322-X}{\emph{Phys. Lett.} {\bf
  160B} (1985) 259--262}.

\bibitem{Alkalaev:2013fsa}
K.~B. Alkalaev, \emph{{On higher spin extension of the Jackiw-Teitelboim
  gravity model}},
  \href{http://dx.doi.org/10.1088/1751-8113/47/36/365401}{\emph{J. Phys.} {\bf
  A47} (2014) 365401}, [\href{http://arxiv.org/abs/1311.5119}{{\tt
  1311.5119}}].

\bibitem{Alkalaev:2014qpa}
K.~B. Alkalaev, \emph{{Global and local properties of AdS$_{2}$ higher spin
  gravity}}, \href{http://dx.doi.org/10.1007/JHEP10(2014)122}{\emph{JHEP} {\bf
  10} (2014) 122}, [\href{http://arxiv.org/abs/1404.5330}{{\tt 1404.5330}}].

\bibitem{Grumiller:2013swa}
D.~Grumiller, M.~Leston and D.~Vassilevich, \emph{{Anti-de Sitter holography
  for gravity and higher spin theories in two dimensions}},
  \href{http://dx.doi.org/10.1103/PhysRevD.89.044001}{\emph{Phys. Rev.} {\bf
  D89} (2014) 044001}, [\href{http://arxiv.org/abs/1311.7413}{{\tt
  1311.7413}}].

\bibitem{Grumiller:2015vaa}
D.~Grumiller, J.~Salzer and D.~Vassilevich, \emph{{AdS$_{2}$ holography is
  (non-)trivial for (non-)constant dilaton}},
  \href{http://dx.doi.org/10.1007/JHEP12(2015)015}{\emph{JHEP} {\bf 12} (2015)
  015}, [\href{http://arxiv.org/abs/1509.08486}{{\tt 1509.08486}}].

\bibitem{Gonzalez:2018enk}
H.~A. González, D.~Grumiller and J.~Salzer, \emph{{Towards a bulk description
  of higher spin SYK}},
  \href{http://dx.doi.org/10.1007/JHEP05(2018)083}{\emph{JHEP} {\bf 05} (2018)
  083}, [\href{http://arxiv.org/abs/1802.01562}{{\tt 1802.01562}}].

\bibitem{Feigin}
B.~L. Feigin, \emph{{Lie algebras $gl(\lambda)$ and cohomologies of Lie
  algebras of differential operators}}, {\emph{Russian Math. Surv.} {\bf 43}
  (1988) 169--170}.

\bibitem{Vasiliev:1989re}
M.~A. Vasiliev, \emph{{Higher Spin Algebras and Quantization on the Sphere and
  Hyperboloid}}, \href{http://dx.doi.org/10.1142/S0217751X91000605}{\emph{Int.
  J. Mod. Phys.} {\bf A6} (1991) 1115--1135}.

\bibitem{Jevicki:2016bwu}
A.~Jevicki, K.~Suzuki and J.~Yoon, \emph{{Bi-Local Holography in the SYK
  Model}}, \href{http://dx.doi.org/10.1007/JHEP07(2016)007}{\emph{JHEP} {\bf
  07} (2016) 007}, [\href{http://arxiv.org/abs/1603.06246}{{\tt 1603.06246}}].

\bibitem{Jevicki:2016ito}
A.~Jevicki and K.~Suzuki, \emph{{Bi-Local Holography in the SYK Model:
  Perturbations}}, \href{http://dx.doi.org/10.1007/JHEP11(2016)046}{\emph{JHEP}
  {\bf 11} (2016) 046}, [\href{http://arxiv.org/abs/1608.07567}{{\tt
  1608.07567}}].

\bibitem{Das:2003vw}
S.~R. Das and A.~Jevicki, \emph{{Large N collective fields and holography}},
  \href{http://dx.doi.org/10.1103/PhysRevD.68.044011}{\emph{Phys. Rev.} {\bf
  D68} (2003) 044011}, [\href{http://arxiv.org/abs/hep-th/0304093}{{\tt
  hep-th/0304093}}].

\bibitem{Gross:2017vhb}
D.~J. Gross and V.~Rosenhaus, \emph{{A line of CFTs: from generalized free
  fields to SYK}}, \href{http://dx.doi.org/10.1007/JHEP07(2017)086}{\emph{JHEP}
  {\bf 07} (2017) 086}, [\href{http://arxiv.org/abs/1706.07015}{{\tt
  1706.07015}}].

\bibitem{Peng:2018zap}
C.~Peng, \emph{{$\mathcal{N}=(0,2)$ SYK, Chaos and Higher-Spins}},
  \href{http://dx.doi.org/10.1007/JHEP12(2018)065}{\emph{JHEP} {\bf 12} (2018)
  065}, [\href{http://arxiv.org/abs/1805.09325}{{\tt 1805.09325}}].

\bibitem{Bekaert:2011js}
X.~Bekaert, \emph{{Singletons and their maximal symmetry algebras}},  in
  \emph{{Modern Mathematical Physics. Proceedings, 6th Summer School: Belgrade,
  Serbia, September 14-23, 2010}}, pp.~71--89, 2011.
\newblock \href{http://arxiv.org/abs/1111.4554}{{\tt 1111.4554}}.

\bibitem{Angelopoulos:1997ij}
E.~Angelopoulos and M.~Laoues, \emph{{Masslessness in n-dimensions}},
  \href{http://dx.doi.org/10.1142/S0129055X98000082}{\emph{Rev. Math. Phys.}
  {\bf 10} (1998) 271--300}, [\href{http://arxiv.org/abs/hep-th/9806100}{{\tt
  hep-th/9806100}}].

\bibitem{Angelopoulos:1980wg}
E.~Angelopoulos, M.~Flato, C.~Fronsdal and D.~Sternheimer, \emph{Massless
  particles, conformal group and de sitter universe},
  \href{http://dx.doi.org/10.1103/PhysRevD.23.1278}{\emph{Phys. Rev.} {\bf D23}
  (1981) 1278}.

\bibitem{Breitenlohner:1982jf}
P.~Breitenlohner and D.~Z. Freedman, \emph{{Stability in Gauged Extended
  Supergravity}},
  \href{http://dx.doi.org/10.1016/0003-4916(82)90116-6}{\emph{Ann. Phys.} {\bf
  144} (1982) 249}.

\bibitem{Kitaev:2017hnr}
A.~Kitaev, \emph{{Notes on $\widetilde{\mathrm{SL}}(2,\mathbb{R})$
  representations}},  \href{http://arxiv.org/abs/1711.08169}{{\tt 1711.08169}}.

\bibitem{Klebanov:1999tb}
I.~R. Klebanov and E.~Witten, \emph{{AdS / CFT correspondence and symmetry
  breaking}},
  \href{http://dx.doi.org/10.1016/S0550-3213(99)00387-9}{\emph{Nucl. Phys.}
  {\bf B556} (1999) 89--114}, [\href{http://arxiv.org/abs/hep-th/9905104}{{\tt
  hep-th/9905104}}].

\bibitem{Gaberdiel:2010pz}
M.~R. Gaberdiel and R.~Gopakumar, \emph{{An AdS Dual for Minimal Model CFTs}},
  \href{http://dx.doi.org/10.1103/PhysRevD.83.066007}{\emph{Phys. Rev.} {\bf
  D83} (2011) 066007}, [\href{http://arxiv.org/abs/1011.2986}{{\tt
  1011.2986}}].

\bibitem{Sezgin:2003pt}
E.~Sezgin and P.~Sundell, \emph{{Holography in 4D (super) higher spin theories
  and a test via cubic scalar couplings}},
  \href{http://dx.doi.org/10.1088/1126-6708/2005/07/044}{\emph{JHEP} {\bf 07}
  (2005) 044}, [\href{http://arxiv.org/abs/hep-th/0305040}{{\tt
  hep-th/0305040}}].

\bibitem{Alkalaev:2012rg}
K.~Alkalaev, \emph{{Mixed-symmetry tensor conserved currents and AdS/CFT
  correspondence}},
  \href{http://dx.doi.org/10.1088/1751-8113/46/21/214007}{\emph{J. Phys.} {\bf
  A46} (2013) 214007}, [\href{http://arxiv.org/abs/1207.1079}{{\tt
  1207.1079}}].

\bibitem{Basile:2018dzi}
T.~Basile, X.~Bekaert and E.~Joung, \emph{{Twisted Flato-Fronsdal Theorem for
  Higher-Spin Algebras}},
  \href{http://dx.doi.org/10.1007/JHEP07(2018)009}{\emph{JHEP} {\bf 07} (2018)
  009}, [\href{http://arxiv.org/abs/1802.03232}{{\tt 1802.03232}}].

\bibitem{Grigoriev:2019xmp}
M.~Grigoriev, I.~Lovrekovic and E.~Skvortsov, \emph{{New Conformal Higher Spin
  Gravities in $3d$}},  \href{http://arxiv.org/abs/1909.13305}{{\tt
  1909.13305}}.

\bibitem{Basile:2014wua}
T.~Basile, X.~Bekaert and N.~Boulanger, \emph{{Flato-Fronsdal theorem for
  higher-order singletons}},
  \href{http://dx.doi.org/10.1007/JHEP11(2014)131}{\emph{JHEP} {\bf 11} (2014)
  131}, [\href{http://arxiv.org/abs/1410.7668}{{\tt 1410.7668}}].

\bibitem{Brust:2016gjy}
C.~Brust and K.~Hinterbichler, \emph{{Free $\Box^{k}$ scalar conformal field
  theory}}, \href{http://dx.doi.org/10.1007/JHEP02(2017)066}{\emph{JHEP} {\bf
  02} (2017) 066}, [\href{http://arxiv.org/abs/1607.07439}{{\tt 1607.07439}}].

\bibitem{Flato:1978qz}
M.~Flato and C.~Fronsdal, \emph{{One Massless Particle Equals Two Dirac
  Singletons: Elementary Particles in a Curved Space. 6}},
  \href{http://dx.doi.org/10.1007/BF00400170}{\emph{Lett. Math. Phys.} {\bf 2}
  (1978) 421--426}.

\bibitem{Vasiliev:2004cm}
M.~A. Vasiliev, \emph{{Higher spin superalgebras in any dimension and their
  representations}},
  \href{http://dx.doi.org/10.1088/1126-6708/2004/12/046}{\emph{JHEP} {\bf 12}
  (2004) 046}, [\href{http://arxiv.org/abs/hep-th/0404124}{{\tt
  hep-th/0404124}}].

\bibitem{Dolan:2005wy}
F.~A. Dolan, \emph{{Character formulae and partition functions in higher
  dimensional conformal field theory}},
  \href{http://dx.doi.org/10.1063/1.2196241}{\emph{J. Math. Phys.} {\bf 47}
  (2006) 062303}, [\href{http://arxiv.org/abs/hep-th/0508031}{{\tt
  hep-th/0508031}}].

\bibitem{Metsaev:2000qb}
R.~R. Metsaev, \emph{{Massive fields in AdS(3) and compactification in AdS
  space time}},
  \href{http://dx.doi.org/10.1016/S0920-5632(01)01543-2}{\emph{Nucl. Phys.
  Proc. Suppl.} {\bf 102} (2001) 100--106},
  [\href{http://arxiv.org/abs/hep-th/0103088}{{\tt hep-th/0103088}}].

\bibitem{Artsukevich:2008vy}
A.~Y. Artsukevich and M.~A. Vasiliev, \emph{{On Dimensional Degression in
  AdS(d)}}, \href{http://dx.doi.org/10.1103/PhysRevD.79.045007}{\emph{Phys.
  Rev.} {\bf D79} (2009) 045007}, [\href{http://arxiv.org/abs/0810.2065}{{\tt
  0810.2065}}].

\bibitem{Gross:2017aos}
D.~J. Gross and V.~Rosenhaus, \emph{{All point correlation functions in SYK}},
  \href{http://dx.doi.org/10.1007/JHEP12(2017)148}{\emph{JHEP} {\bf 12} (2017)
  148}, [\href{http://arxiv.org/abs/1710.08113}{{\tt 1710.08113}}].

\bibitem{Prokushkin:1998bq}
S.~F. Prokushkin and M.~A. Vasiliev, \emph{{Higher spin gauge interactions for
  massive matter fields in 3-D AdS space-time}},
  \href{http://dx.doi.org/10.1016/S0550-3213(98)00839-6}{\emph{Nucl. Phys.}
  {\bf B545} (1999) 385}, [\href{http://arxiv.org/abs/hep-th/9806236}{{\tt
  hep-th/9806236}}].

\bibitem{Konstein:1989ij}
S.~E. Konstein and M.~A. Vasiliev, \emph{Extended higher spin superalgebras and
  their massless representations},
  \href{http://dx.doi.org/10.1016/0550-3213(90)90216-Z}{\emph{Nucl. Phys.} {\bf
  B331} (1990) 475--499}.

\bibitem{Fradkin:1990ki}
E.~S. Fradkin and V.~{\relax Ya}. Linetsky, \emph{{Infinite dimensional
  generalizations of finite dimensional symmetries}},
  \href{http://dx.doi.org/10.1063/1.529318}{\emph{J. Math. Phys.} {\bf 32}
  (1991) 1218--1226}.

\bibitem{Fradkin:1990ir}
E.~S. Fradkin and V.~{\relax Ya}. Linetsky, \emph{{Infinite dimensional
  generalizations of simple Lie algebras}},
  \href{http://dx.doi.org/10.1142/S0217732390002249}{\emph{Mod. Phys. Lett.}
  {\bf A5} (1990) 1967--1977}.

\bibitem{Boulanger:2013zza}
N.~Boulanger, D.~Ponomarev, E.~D. Skvortsov and M.~Taronna, \emph{{On the
  uniqueness of higher-spin symmetries in AdS and CFT}},
  \href{http://dx.doi.org/10.1142/S0217751X13501625}{\emph{Int. J. Mod. Phys.}
  {\bf A28} (2013) 1350162}, [\href{http://arxiv.org/abs/1305.5180}{{\tt
  1305.5180}}].

\bibitem{Joung:2014qya}
E.~Joung and K.~Mkrtchyan, \emph{{Notes on higher-spin algebras: minimal
  representations and structure constants}},
  \href{http://dx.doi.org/10.1007/JHEP05(2014)103}{\emph{JHEP} {\bf 05} (2014)
  103}, [\href{http://arxiv.org/abs/1401.7977}{{\tt 1401.7977}}].

\bibitem{Fernando:2015tiu}
S.~Fernando and M.~Gunaydin, \emph{{Massless conformal fields,
  $AdS_{d+1}/CFT_d$ higher spin algebras and their deformations}},
  \href{http://dx.doi.org/10.1016/j.nuclphysb.2016.01.024}{\emph{Nucl. Phys.}
  {\bf B904} (2016) 494--526}, [\href{http://arxiv.org/abs/1511.02167}{{\tt
  1511.02167}}].

\bibitem{Vasiliev:2003ev}
M.~A. Vasiliev, \emph{{N}onlinear equations for symmetric massless higher spin
  fields in ({A})d{S}(d)}, {\emph{Phys. Lett.} {\bf B567} (2003) 139--151},
  [\href{http://arxiv.org/abs/hep-th/0304049}{{\tt hep-th/0304049}}].

\bibitem{Howe}
R.~Howe, \emph{Transcending classical invariant theory}, {\emph{J. Amer. Math.
  Soc.} {\bf 3} (1989) 2}.

\bibitem{Vasiliev:1999ba}
M.~A. Vasiliev, \emph{Higher spin gauge theories: Star-product and {AdS}
  space},  \href{http://arxiv.org/abs/hep-th/9910096}{{\tt hep-th/9910096}}.

\bibitem{Arnaudon}
D.~Arnaudon, M.~Bauer and L.~Frappat, \emph{{On Casimir's Ghost}},
  \href{http://dx.doi.org/10.1007/s002200050143}{\emph{Commun.Math.Phys.} {\bf
  187} (1997) 429}, [\href{http://arxiv.org/abs/q-alg/9605021}{{\tt
  q-alg/9605021}}].

\bibitem{Iazeolla:2008ix}
C.~Iazeolla and P.~Sundell, \emph{{A Fiber Approach to Harmonic Analysis of
  Unfolded Higher-Spin Field Equations}},
  \href{http://dx.doi.org/10.1088/1126-6708/2008/10/022}{\emph{JHEP} {\bf 10}
  (2008) 022}, [\href{http://arxiv.org/abs/0806.1942}{{\tt 0806.1942}}].

\bibitem{Stelle:1979aj}
K.~S. Stelle and P.~C. West, \emph{Spontaneously broken {D}e {S}itter symmetry
  and the gravitational holonomy group}, {\emph{Phys. Rev.} {\bf D21} (1980)
  1466}.

\bibitem{Bekaert:2005vh}
X.~Bekaert, S.~Cnockaert, C.~Iazeolla and M.~A. Vasiliev, \emph{{Nonlinear
  higher spin theories in various dimensions}},
  \href{http://arxiv.org/abs/hep-th/0503128}{{\tt hep-th/0503128}}.

\bibitem{Bekaert:2005in}
X.~Bekaert and J.~Mourad, \emph{{The Continuous spin limit of higher spin field
  equations}},
  \href{http://dx.doi.org/10.1088/1126-6708/2006/01/115}{\emph{JHEP} {\bf 01}
  (2006) 115}, [\href{http://arxiv.org/abs/hep-th/0509092}{{\tt
  hep-th/0509092}}].

\bibitem{Alkalaev:2017hvj}
K.~B. Alkalaev and M.~A. Grigoriev, \emph{{Continuous spin fields of
  mixed-symmetry type}},
  \href{http://dx.doi.org/10.1007/JHEP03(2018)030}{\emph{JHEP} {\bf 03} (2018)
  030}, [\href{http://arxiv.org/abs/1712.02317}{{\tt 1712.02317}}].

\bibitem{Bekaert:2017xin}
X.~Bekaert, J.~Mourad and M.~Najafizadeh, \emph{{Continuous-spin field
  propagator and interaction with matter}},
  \href{http://arxiv.org/abs/1710.05788}{{\tt 1710.05788}}.

\bibitem{Alkalaev:2018bqe}
K.~Alkalaev, A.~Chekmenev and M.~Grigoriev, \emph{{Unified formulation for
  helicity and continuous spin fermionic fields}},
  \href{http://dx.doi.org/10.1007/JHEP11(2018)050}{\emph{JHEP} {\bf 11} (2018)
  050}, [\href{http://arxiv.org/abs/1808.09385}{{\tt 1808.09385}}].

\bibitem{Vasiliev:1994gr}
M.~A. Vasiliev, \emph{Unfolded representation for relativistic equations in
  (2+1) anti-{D}e {S}itter space}, {\emph{Class. Quant. Grav.} {\bf 11} (1994)
  649--664}.

\bibitem{Vasiliev:1995sv}
M.~A. Vasiliev, \emph{{Higher spin gauge interactions for matter fields in
  two-dimensions}},
  \href{http://dx.doi.org/10.1016/0370-2693(95)01122-7}{\emph{Phys. Lett.} {\bf
  B363} (1995) 51--57}, [\href{http://arxiv.org/abs/hep-th/9511063}{{\tt
  hep-th/9511063}}].

\bibitem{Shaynkman:2000ts}
O.~V. Shaynkman and M.~A. Vasiliev, \emph{{Scalar field in any dimension from
  the higher spin gauge theory perspective}},
  \href{http://dx.doi.org/10.1007/BF02551402}{\emph{Theor. Math. Phys.} {\bf
  123} (2000) 683--700}, [\href{http://arxiv.org/abs/hep-th/0003123}{{\tt
  hep-th/0003123}}].

\bibitem{Vasiliev:1990vu}
M.~A. Vasiliev, \emph{{Properties of equations of motion of interacting gauge
  fields of all spins in (3+1)-dimensions}},
  \href{http://dx.doi.org/10.1088/0264-9381/8/7/014}{\emph{Class. Quant. Grav.}
  {\bf 8} (1991) 1387--1417}.

\bibitem{Vasiliev:1992av}
M.~A. Vasiliev, \emph{More on equations of motion for interacting massless
  fields of all spins in (3+1)-dimensions}, {\emph{Phys. Lett.} {\bf B285}
  (1992) 225--234}.

\bibitem{Alkalaev:2014nsa}
K.~B. Alkalaev, M.~Grigoriev and E.~D. Skvortsov, \emph{{Uniformizing
  higher-spin equations}},
  \href{http://dx.doi.org/10.1088/1751-8113/48/1/015401}{\emph{J. Phys.} {\bf
  A48} (2015) 015401}, [\href{http://arxiv.org/abs/1409.6507}{{\tt
  1409.6507}}].

\bibitem{Vasiliev:2018zer}
M.~A. Vasiliev, \emph{{From Coxeter Higher-Spin Theories to Strings and Tensor
  Models}}, \href{http://dx.doi.org/10.1007/JHEP08(2018)051}{\emph{JHEP} {\bf
  08} (2018) 051}, [\href{http://arxiv.org/abs/1804.06520}{{\tt 1804.06520}}].

\bibitem{Sharapov:2019vyd}
A.~Sharapov and E.~Skvortsov, \emph{{Formal Higher Spin Gravities}},
  \href{http://dx.doi.org/10.1016/j.nuclphysb.2019.02.011}{\emph{Nucl. Phys.}
  {\bf B941} (2019) 838--860}, [\href{http://arxiv.org/abs/1901.01426}{{\tt
  1901.01426}}].

\bibitem{Mazorchuk}
V.~Mazorchuk, \emph{Lectures on $sl_2(C)$-modules}.
\newblock Imperial College Press, 2009.

\end{thebibliography}

\providecommand{\href}[2]{#2}\begingroup\raggedright\endgroup

\end{document}